# Fabricating Multifunctional Nanoparticle Membranes by a Fast Layer-by-Layer Langmuir-Blodgett Process: Application in Lithium-sulfur Batteries

*Mun Sek Kim[1,3,†], Lin Ma[2,†], Snehashis Choudhury[1], Surya S. Moganty[4], Shuya Wei[1], and Lynden A. Archer[1,*]*

[1]Department of Chemical and Biomoleular Engineering and [2]Department of Materials Science & Engineering, Cornell University, Ithaca, New York 14853-5201, United States.

[3]Center for Energy Convergence, Korea Institute of Science and Technology, Hwarangno 14-gil 5, Seongbuk-gu, Seoul, 02792, Republic of Korea.

[4]NOHMs Technologies, 1200 Ridgeway Ave. Suite 110, Rochester, NY, 14615, United States.

[†]These authors contributed equally to this work.

[*]**Tel: 607 254-8825. E-mail**: laa25@cornell.edu

**Abstract**

The Langmuir-Blodgett (LB) technique is a powerful, widely used method for preparing coatings of amphiphilic molecules at air/water interfaces with thickness control down to a single molecule. Here we report two new LB techniques designed to create ordered, multifunctional nanoparticle films on any non-reactive support. The methods utilize Marangoni stresses produced by surfactants at a fluid/solid/gas interface and self-assembly of nanoparticles to facilitate rapid creation of dense monolayers of multi-wall carbon nanotubes (MWCNT), metal-oxide nanoparticles, polymers, and combinations of these materials in a layer-by-layer configuration. Using the polyolefin separator in a lithium sulfur (Li-S) electrochemical cell as an example, we illustrate how the method can be used to create structured membranes for regulating mass and charge transport. We show that a layered MWCNT/SiO$_2$/MWCNT nanomaterial created in a clip-like configuration, with gravimetric areal coverage of ~130 μg cm$^{-2}$ and a thickness of ~3 μm, efficiently adsorbs dissolved lithium polysulfide (LiPS) species and efficiently reutilize them for improving Li-S battery performance.

**Introduction**

The Langmuir-Blodgett (LB) technique is a method for preparing coatings of amphiphilic molecules at air/water interfaces with thickness of one molecule[1–3]. The method is attractive for a variety of reasons, including its ability to precisely control the thicknesses of coatings down to molecular dimensions, for the versatility of substrates that can be coated, and for its scalability. Ever since its discovery in the 1920s and its rise in popularity following Irving Langmuir's receipt of the Nobel prize in chemistry in 1932, it has been applied in numerous fields of science and technology to form thin micro patterns,[4,5] molecular-thick thin films,[6,7] and more recently monoparticle layers based on spheres,[8,9] rods,[10,11] and nanotubes,[12] which can be easily transferred onto various substrates. Recently, Nie et al proposed an electrospray method that extends the LB technique to yield high coverage of metallic nanospheres at the surface of water[13]. With LB assembly it is therefore now possible to produce monolayers of colloidal films that broadly range in particle sizes and shapes, which may be used to advantage for tuning film properties[14] as well as for coating applications to create thin film devices[15,16].

This article reports two new and versatile LB coating approaches - Langmuir-Blodgett sequential dip coating (LBSDC) and the Langmuir-Blodgett scooping (LBS), which facilitate efficient creation of multifunctional, layer-by-layer coatings of carbon, metal-oxides, polymers, and combinations of these materials on any non-reactive substrate. Unlike the conventional LB method, which uses a mechanical force applied to a disordered material at the air/water interface to create well-ordered assemblies of molecules or particles, LBSDC and LBS utilize surfactant and self-assembly, respectively, to create ordered coatings that can be transferred to a solid or porous support. This difference allows highly organized coatings to be created in a fraction of the time and using any containment vessel (i.e. a LB mechanical barrier is not needed). The speed with which ordered monolayer coatings are created, the high quality and low thickness of the transferred coatings, and versatility of the process by which the coatings are formed mean that LBSDC and LBS can be applied in a repetitive fashion to create multi-functional coatings in a layer-by-layer format that enable design of new materials with surface features able to regulate mass and charge transport. Because the assembly occurs at a sharp gas/liquid interface, the methods nonetheless benefit from the inherent attributes of the LB technique - precise control over film thickness and structure, as well as the versatility of substrate choices. Moreover, numbers and positions of suspension injection nozzle and water surface area can be altered and customized to scale up the coating process.

The utility of the LBSDC and LBS approaches is illustrated in the present study using the polyolefin separator membrane of a standard Lithium-Sulfur (Li-S) electrochemical cell as a substrate. This choice is motivated by the promise such cells offer for cost-effective storage of large quantities of electrical energy and by the stubborn challenges associated with solubility and diffusion of long-chain ($Li_2S_x$; x ≥4), lithium polysulfide (LiPS) species, to the electrolyte that limit performance of Li-S batteries.[17-48] We report that using LBSDC and LBS it is possible to create multifunctional coatings in multiple designs that enable conventional membranes to overcome the most difficult challenges. We further report a novel "clip" separator membrane configuration in which a well-formed, but incomplete layer of structures of one chemistry is sandwiched between complete layers of another chemistry. This coating morphology allows one to engineer the surface of a membrane to simultaneously trap an undesired material (e.g. LiPS) and to maintain electrochemical access to it. In so doing, we show that it is possible to preserve the favorable attributes of the Li-S cell and address some of its most serious weaknesses.

**Results and Discussions**

The LBSDC and LBS coating methods enable creation of well-defined layers of materials in various physical forms and chemistries on a conventional polypropylene separator, without the need for chemical binders. LBSDC is a discontinuous process that utilizes a sodium dodecyl sulfate (SDS) surfactant, inducing Marangoni effect, to lower the surface tension of water and to provide a unidirectional force on floating particles or to a particulate LB film at the air-water interface to form a dense, close-packed structure. Care is needed in this step, for on the small length scales of these monolayer films the pressure provided by the surfactant can easily exceed the stability of the self-assembled LB film, causing it to rupture due to too strong surface tension gradient (Supplementary Figure S1). Grains that exceed 200 nm in size exhibit the greatest film stability and are able to form the most densely packed coatings through the LBSDC technique. The LBS method, on the other hand, is a continuous process that uses constant injection of a particle suspension during the coating process to maintain a closely-packed LB film by a self-assembly mechanism, which is induced by the simple spreading and mixing of the water miscible fluid. This approach enables particles smaller than 200 nm in size to be coated on a mobile substrate due to the absence of the extra surface tension gradient provided by the surfactant and may be implemented in a roll-to-roll manufacturing process. The LBS method is therefore more flexible than the LBSDC method, but requires constant injection of the suspension during the coating process.

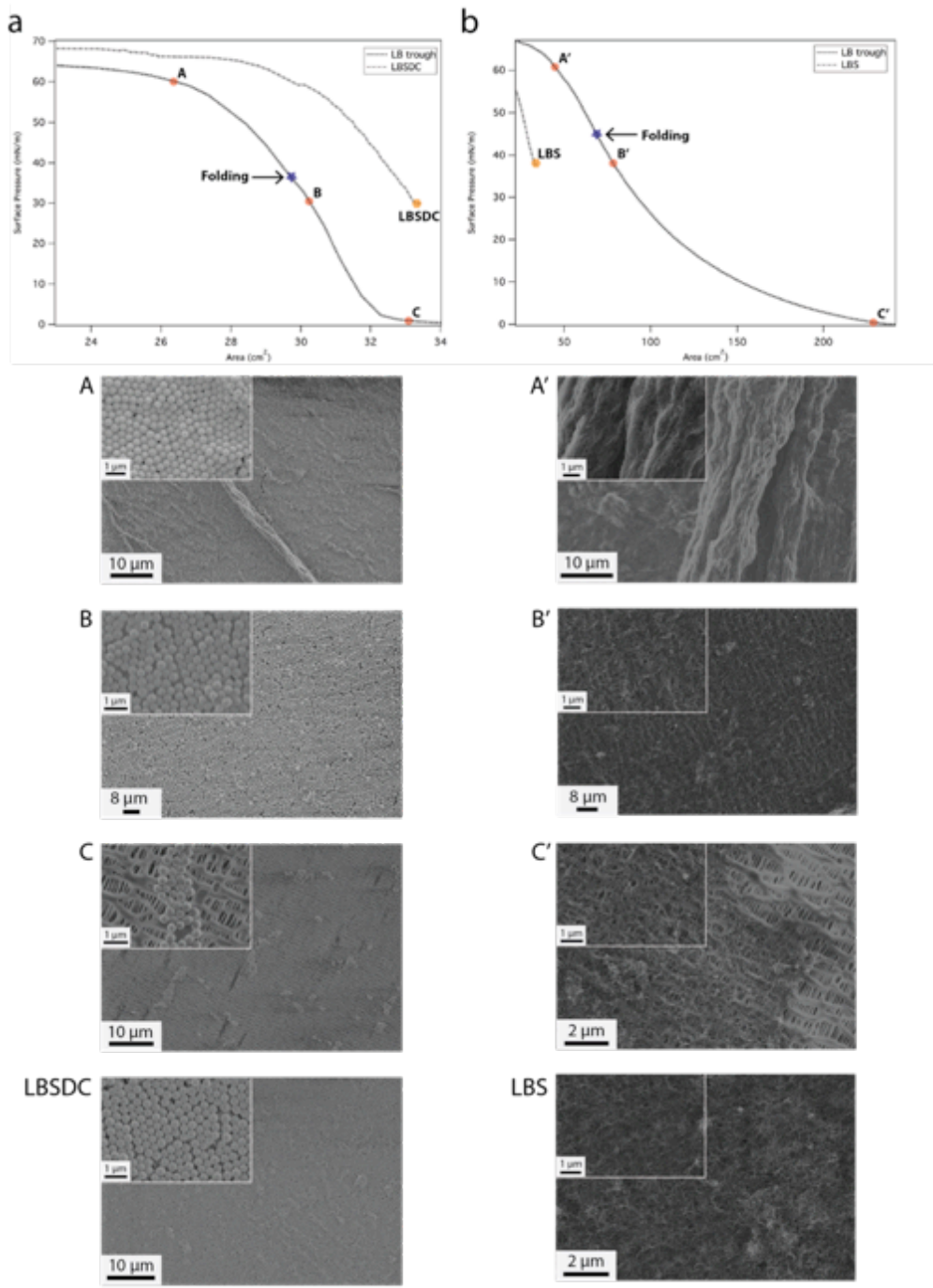

**Figure 1 | Langmuir-Blodgett surface pressure profiles and corresponding coating qualities of silica nanospheres and MWCNTs at designated surface pressures. a**, Silica nanosphere surface pressure profiles of conventional LB trough and LBSDC methods with SEM images of the coating qualities at 60 mN m$^{-1}$, 33 mN m$^{-1}$, and 2 mN m$^{-1}$. **b**, MWCNT surface pressure profiles of conventional LB trough and LBS methods with SEM images of the coating qualities at 61 mN m$^{-1}$, 37 mN m$^{-1}$, and 1 mN m$^{-1}$.

Surface pressure profiles obtained using the LBSDC and LBS approaches to assemble monoparticle layers of ~350 nm diameter sized nanospheres and multi-walled carbon nanotubes (MWCNT) at the air/water interface are reported in Supplementary Figure S2. Figure 1 compares the pressure profiles to those obtained using a conventional LB trough. Three different surface pressure points are chosen in Figure 1a to investigate the packing densities of the colloidal film onto a separator achieved with each of the approaches. The points A, B, and C correspond to the surface pressure of 60 mN m$^{-1}$, 33 mN m$^{-1}$, and 2 mN m$^{-1}$, respectively: Points A and C represent the surface pressure where the colloids are overly packed and inadequately packed, and point B represents the starting surface pressure from the LBSDC method. The inflection point of the surface pressure profile represents a transition point where the folding of the film starts, which is consistent with what is observed in the SEM images at point A. Point B shows the most uniform coating, which indicates that the surface pressure between 33 mN m$^{-1}$ and 38 mN m$^{-1}$ will yield the highest quality LB film. The surface pressure profile of the colloids from the LBSDC method starts at 33 mN m$^{-1}$, which represents the amount of pressure exerted by one 5μL drop of 3wt% SDS surfactant on the film. This pressure from the surface tension gradient allows the colloids to be packed closely and remain stationary, and therefore, no inflection point is observed from the LBSDC profile. To confirm the packing density of the colloids using the LBSDC method, the colloids are coated onto the separator using the LB trough at 33 mN m$^{-1}$ and using the LBSDC method. A coating quality of good agreement is observed from SEM images at point B and from LBSDC, which confirms that LBSDC starts from a highly packed colloidal LB film. To understand the role of the surfactant in the LBSDC method, the maximum amount of the pressure acting on the film is measured and its stability is observed (See Supplementary Figure S3a & S3b). The maximum pressure that the surfactant can provide is ~34 mN m$^{-1}$ and tends to slowly fade over time. The maximum pressure exerted by the surfactant matches the starting surface pressure of the colloids using the LBSDC method, where the increased pressure is the same as obtained from one drop of the surfactant.

In Figure 1b, the surface pressure profiles of the MWCNT is compared from conventional LB trough and LBS methods, and three different surface pressure points, A', B', and C', are chosen at 61 mN m$^{-1}$, 37 mN m$^{-1}$, and 1 mN m$^{-1}$, respectively, to observe the coating quality. The MWCNT film tends to fold as shown in the SEM at point A', and poor coverage of the film is seen at point C'. A long compression region of the film is observed from the profile by comparing the area before and after the inflection point. This is because of the elastic behavior of the self-assembled MWCNT film as the film is comprised of nanotubes. Based on the geometry of the particle, different trends of the surface pressure profiles can be obtained (See Supplementary Figure S3c).

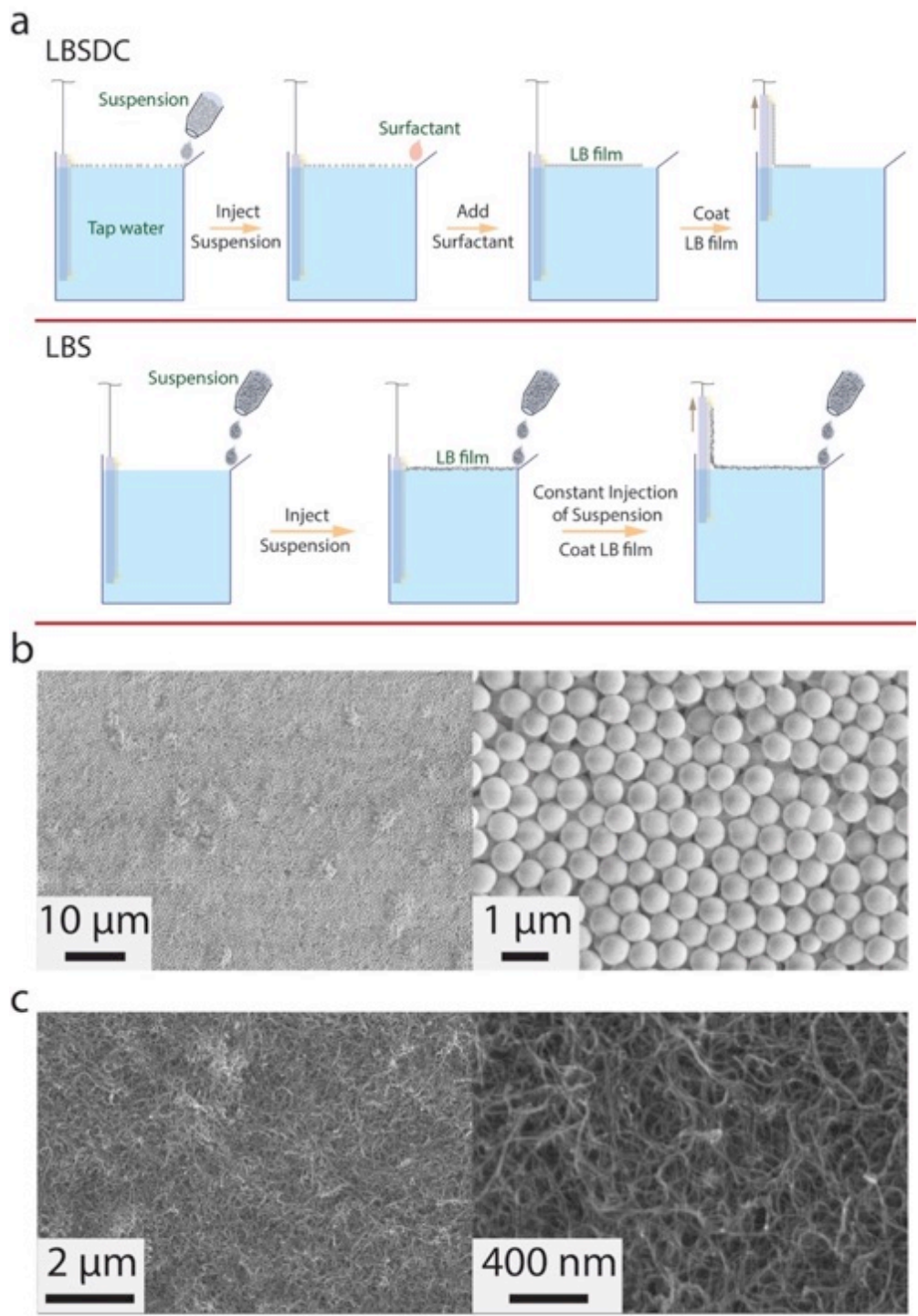

**Figure 2 | LBSDC and LBS coating method schematics and SEM images of multilayer coatings of silica nanospheres and MWCNTs on polypropylene separators. a**, Schematic illustrations of LBSDC and LBS coating methods. **b**, SEM images of three monolayers of silica nanospheres coated separator using LBSDC and **c**, three coating layers of MWCNT coated separator using LBS.

To investigate where the coating quality of the LBS lies, the surface pressure profile of the LBS method is measured and compared to the profile obtained from the LB trough. The LBS method requires a constant injection of the suspension to maintain high packing density by self-assembly mechanism induced by spreading and mixing of the suspension solvent (ethanol) with water. To observe how much pressure is exerted during the self-assembly, the surface pressure was measured by saturating the surface of the trough with MWCNT. The measured pressure is 37 mN m$^{-1}$, which is the pressure exerted from the self-assembly. Since the spreading velocity depends on the distance traveled by the particle, the area on the trough is set to around 25 cm$^2$, which is a similar surface area for our experimental coating process. No inflection point is shown from the profile, which confirms that the fibers are closely packed and compressed from the starting point. Furthermore, congruent coating qualities are observed at point B' and from the LBS in Figure 1b, confirming that the LBS method yields a closely packed, high quality LB film. Moreover, surface pressure profiles of Ketjen Black (KB) and Super P (SP) carbons using the LB trough and LBS methods are measured to understand the stability of the film in the presence of the surfactant (See Supplementary Figure S3c). The starting surface pressure of MWCNT, KB, and SP from the LBS method is 37 mN m$^{-1}$, 20 mN m$^{-1}$, and 35 mN m$^{-1}$, respectively. One drop of the surfactant provides an instant pressure of 34 mN m$^{-1}$, similar to that of MWCNT and SP, while exceeding that of KB. As a result, the film starts to collapse in the presence of the extra surface tension gradient. As expected, the instant destruction of the KB film, ~1 second, is observed, while a longer destruction time, ~7 seconds, is observed for MWCNT and SP in the presence of the surfactant (Supplementary Figure S1). The above results support that the LBSDC and LBS coating methods start in an optimized packing condition and yield high-quality LB films. Figure 2a illustrates the simplicity and effectiveness of the LBSDC and LBS methods. Multilayer coatings using the two approaches are reported in Figure 2b & 2c. The coatings and their processes are important not only because they exhibit an excellent close-packed morphology, but also because they are the thinnest and highest fidelity coatings reported on a battery separator. For example, thickness variations for silica nanospheres and MWCNT within a single monolayer of the silica particle size and ~80 nm of MWCNT (Supplementary Figure S4a) are achieved. This means that these coatings will add very little mass to a battery separator or electrode, yet significantly optimize the electrochemical performance in the batteries due to the uniform and densely-packed coating layers.

The polypropylene Celgard$^{TM}$ membrane used as a separator in the Li-S battery is chosen as a substrate to illustrate the utility of the LBSDC and LBS approach for at least three reasons. First, the rechargeable Li-S battery is arguably one of the most important platforms for storing large amounts of electrical energy at a moderate cost. The redox reaction between lithium and sulfur, $16Li + S_8 \Leftrightarrow 8Li_2S$, occurs spontaneously, is reversible, and produces up to two electrons per formula unit of sulfur, without intervention with catalysts or other means. These features endow the Li-S battery with high theoretical specific energy, 2600 Wh kg$^{-1}$, and low material and operating costs[17–19].

Second, in practice Li-S cells fail to deliver on these high expectations for two stubborn, fundamental reasons: **(i)** sulfur and its reduction compounds with lithium are such poor conductors that unless the electrochemical reactions between Li$^+$ and sulfur occur in solution near a conductive substrate or in subnanometer-sized pores of a conductive host material such as microporous carbon, only a small fraction of the active sulfur material in the cathode is electrochemically accessible; and **(ii)** the reaction between Li$^+$ and S$_8$ is a multi-step reaction[20], in which the higher molecular weight intermediate species $Li_2S_x$ (x ≥4), collectively termed lithium polysulfides (LiPS), are soluble whereas the lower molecular weight ones (x < 3) are not. Dissolution of LiPS in an electrolyte means that a substantial fraction of the active material can be lost before it is fully reduced to Li$_2$S, if the LiPS diffuses too far from the conductive substrate in the cathode. An even greater concern is that once in the electrolyte, LiPS can diffuse to the lithium metal anode and undergo chemical reduction to form polysulfides of lower order, some of which are insoluble and deposit on the anode, causing time-dependent loss of both lithium and sulfur in a parasitic process termed shuttling.

Finally, a variety of approaches have been investigated for controlling LiPS dissolution, diffusion, and shuttling in Li-S cells. Methods ranging from sequestering the sulfur in porous carbon structures in nanospheres[21], nanotubes/nanofibers[22], graphene/graphene oxide sheets[23,24] all utilize the strong affinity of sulfur for carbon-based materials to limit dissolution. Other workers have shown that strong specific interactions of LiPS with amine-containing molecular[25,26] and inorganic chalcogenide[27], particulate additives can be used to reduce sulfur loss to the electrolyte[28,29]. Even in the best cases, however, there is a finite, equilibrium concentration of LiPS dissolved in the electrolyte such that chemical potential of LiPS in the cathode is equal to that in the electrolyte[18,25]. As a result, the dissolved LiPS are still able to diffuse to the Li anode, react with it, and increase the interfacial resistance of the anode. Very recently, Hendrickson et al showed that a substantial amount of LiPS is also lost by adsorption in the pores of the separator and that this source of loss can be removed in model Li-S cells run in a separator-/membrane-free configuration, but at the price of very high interfacial impedances at the anode[30]. Other works have shown that incorporation of carbon,[31–42] metal-oxide,[43,44] and polymer[45–47] coatings on separators can reduce LiPS loss, but the electrolyte must still be reinforced with additives such as $LiNO_3$ thought to limit LiPS reaction with metallic lithium, for stable cell cycling over extended periods or in practical lithium- and electrolyte-lean Li-S cell designs.

A broad range of materials such as SP carbon[36–38], KB carbon[37], carbon nanofibers/tubes[32-35], mesoporous carbon[31], alumina[48] and graphene[39-43] were coated on Celgard, and the electrochemical performances of Li-S cells based on these separators are investigated in literatures. To note the versatility and adaptability of the developed coating methods, large selections of a material with one or more different coating materials (Supplementary Figure S5) are coated on the separator and are suitable for different substrates (Supplementary Figure S6 & S7). The thickness of a single layer coating of MWCNT, KB, and SP is ~80 nm, ~350 nm, and ~850 nm, respectively (Supplementary Figure S4). The corresponding gravimetric coverage of a single layer of MWCNT, KB, SP, and ~350 nm silica nanospheres is ~5 µg cm$^{-2}$, ~17 µg cm$^{-2}$, ~20 µg cm$^{-2}$, and ~25 µg cm$^{-2}$, respectively (Supplementary Figure S8). The negligible weight gained per coating layer with high uniformity is best appreciated by comparison to literature results, where carbon materials are coated using the vacuum filtration[32-34] (Loading: 0.17 to 0.35 mg cm$^{-2}$, Thickness: 20 to 25µm) or doctor-blade[31,36-37] (Loading: 0.26 to 0.53 mg cm$^{-2}$, Thickness: 6.7 to 27 µm) methods. As illustrated in Supplementary Figure S9, LBS coatings on Celgard are single-sided and exhibit high mechanical strength absence of chemical binders (Supplementary Figure S10). The effectiveness of LBS-coated Celgard comprised of 1-10 coating layers of silica nanospheres, MWCNT, KB, and SP were systematically studied for their ability to improve cycling behavior in Li-S cells. As shown in Supplementary Figure S11, Li-S cells based on the carbon coated separator yield superior capacity and retention rates, compared with pristine Celgard separator. Specifically, the capacity retention after 100 cycles is improved from 31% for the pristine separator to 63%, 71%, 63%, and 49% for ten coating layers of MWCNT, KB, SP, and silica nanospheres, respectively. The initial capacity for the pristine separator, 10LR MWCNT, 10LR KB, 10LR SP, and 10LR are 1067 mAh g$^{-1}$, 1535 mAh g$^{-1}$, 1594 mAh g$^{-1}$, 1541 mAh g$^{-1}$, and 1588 mAh g$^{-1}$, respectively at 0.5C for the first four and 0.2C for the last.

Our results show that carbon-coated Celgard is far more effective than the silica-coated material in stabilizing cycling performance of Li-S cells. We attribute this behavior to the stronger adsorption of LiPS on the SiO$_2$ coating layer by physical and covalent bonds[28] and the inability to utilize the trapped LiPS. Our results also confirm observations reported by Yao et al[37], that MWCNT and KB are particularly effective as separator coatings because the interconnected porous structure of coatings based on these carbon materials allow for both trapping of LiPS and utilization of the trapped materials in electrochemical cycling. It is important to note, however, that the weight of MWCNT per coating layer is only 25% that of KB, implying that the MWCNT coating is by far the most efficient of the carbon materials studied.

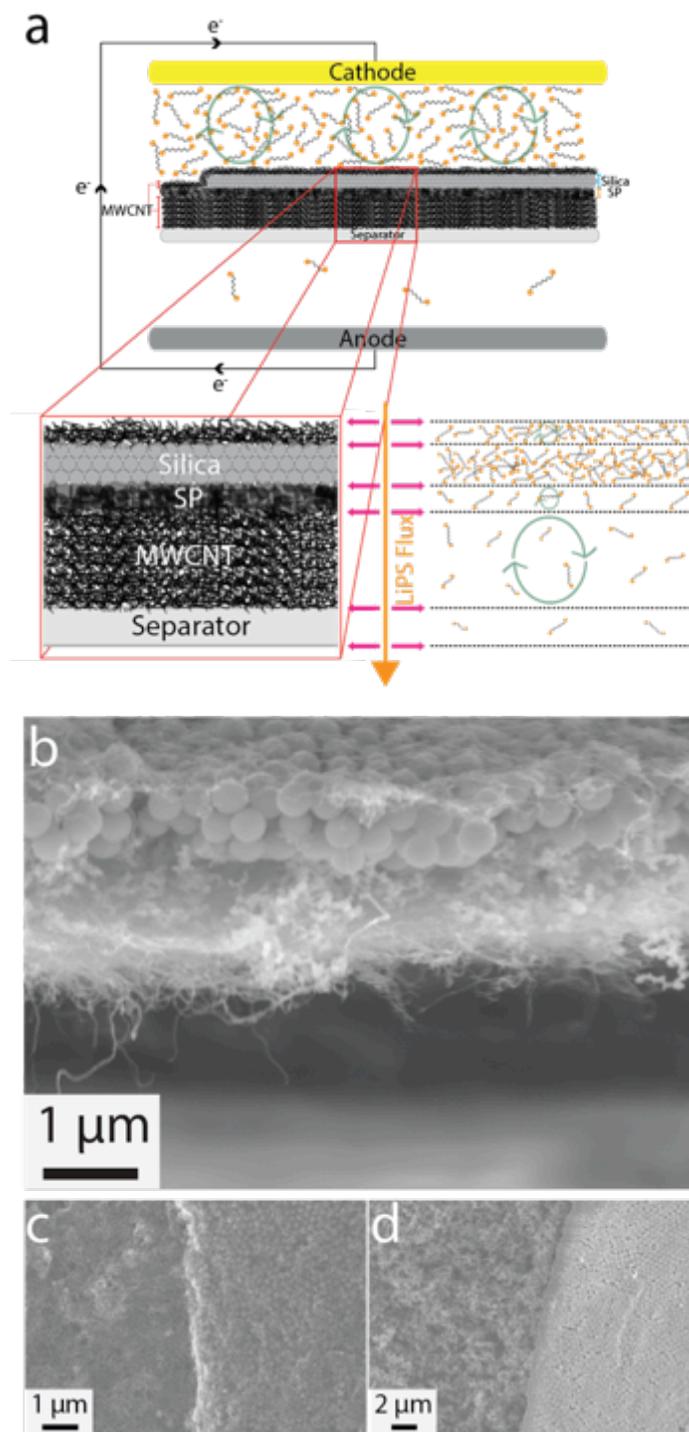

**Figure 3 | Clip configuration coating schematic illustration and SEM images of the clip coated separator. a**, Clip configuration architecture design and LiPS flux diagram across the separator during the discharge of Li-S cell. **b**, Cross-sectional SEM image of the clip coated separator. **c**, SEM image of the clip coated separator at the silica-SP boundary layer. **d**, SEM image of the clip coated separator at the boundary without final MWCNT coating.

A separator coating design that offers a combination of the strong LiPS binding attributes of a close-packed array of $SiO_2$ particles and high utilization of trapped LiPS evident for MWCNT would seem ideal for Li-S cells. This perspective is at odds with the work of Yao et al[37], which previously demonstrated that a Li-S battery separator coated with a mixture of ceramic nanoparticles and Super-P (SP) carbon, using the doctor-blade coating method, yields cells with poorer electrochemical performance than those in which a simple SP coating layer was used. Here, we take advantage of the spatial control afforded by the LBS and LBSDC coating strategy to create a multifunctional separator coating with the configuration shown in Figure 3a. In this so-called *clip* configuration multiple layers of closely packed silica particles are surrounded by a conductive fibrous network based on MWCNT. The location of the silica layer is designed such that under compression in a battery, the two MWCNT coatings contact each other (like the clasps of a clip) and also make contact with the cathode so as to ensure maximum electrochemical access to LiPS trapped in any of the coating layers that comprise the clip. As a proof of concept, we created and studied clip coating designs comprised of five coating layers of MWCNT and three monolayers of silica. The quality and mechanical strength of these coatings are illustrated in Supplementary Figure S12. Figure 3b shows the cross section of the clip configuration, where it is seen that the material has a consistent structure and a thickness of ~3μm. The clip configuration of the coating has also been confirmed by SEM image at the silica-carbon layer boundary (Figure 3c). Figure 3c then shows a uniform thin fibrous morphology of the coating surface after the final layer of MWCNT coverage is established over the three layers of silica nanosphere film (Zoomed-in SEM images is shown in Supplementary Figure S13). We observed the immersion of close-packed silica layers at the third layer on top of the SP layer of the clip coating (Figure 3d). For reference, one monolayer of silica nanospheres is also shown in Supplementary Figure S14.

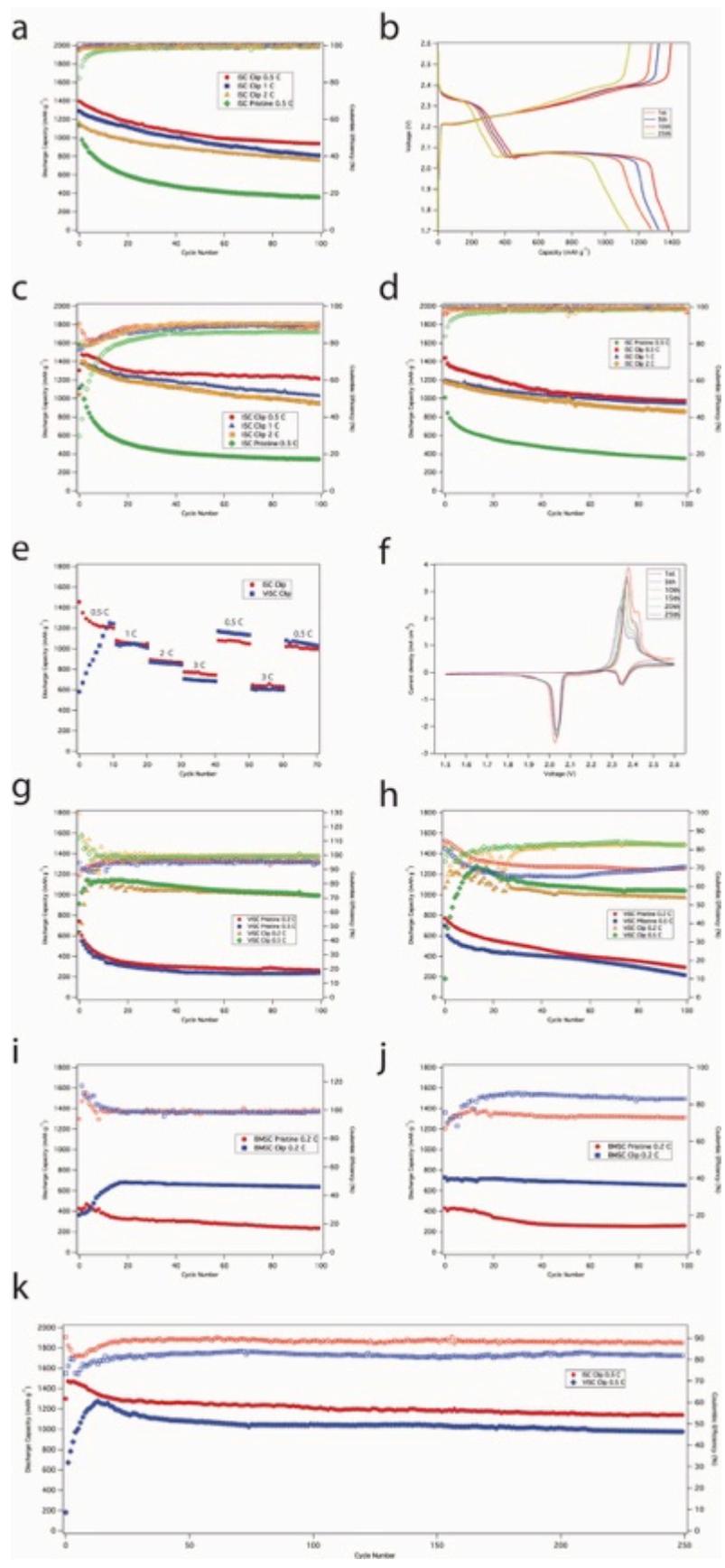

**Figure 4 | Electrochemical performance of the clip coated separator Li-S cells with various cathodes. a**, Cycling performance of the pretreated Li anode Li-S cell with/without the clip coated separator and with ISC at three different C rates. **b**, Discharge-charge voltage profiles of the pretreated Li anode Li-S cell with the clip coated separator and with ISC at 0.5 C. **c**, Cycling performance of pristine Li anode of the Li-S cell with/without the clip coated separator and with ISC at three different C rates without $LiNO_3$ in the electrolyte. **d**, Cycling performance of pristine Li anode of the Li-S cell with/without the clip coated separator and with ISC at three different C rates with 0.05M $LiNO_3$ in the electrolyte. **e**, Cycling performance of pristine Li anode with the clip coated separator Li-S cells with ISC and VISC at various C rates without $LiNO_3$ in the electrolyte. **f**, Cyclic voltammograms of the pristine Li anode with the clip coated separator Li-S cell with ISC at 0.1 mV s$^{-1}$ for various cycles. **g**, Cycling performance of the pristine Li anode Li-S cells with/without the clip coated separator, with VISC, and 0.3M $LiNO_3$ in the electrolyte at two different C rates. **h**, Cycling performance of the pristine Li anode Li-S cells with/without the clip coated separator and with VISC at two different C rates without $LiNO_3$ in the electrolyte. **i**, Cycling performance of the pristine Li anode Li-S cells with/without the clip coated separator, with BMSC, and 0.3M $LiNO_3$ in the electrolyte at 0.2 C. **j**, Cycling performance of the pristine Li anode Li-S cells with/without the clip coated separator and with BMSC at 0.2 C without $LiNO_3$ in the electrolyte. **k**, Cycling performance of the clip coated separator Li-S cells no $LiNO_3$ in the electrolyte and with ISC and VISC at 0.5 C for 250 cycles.

In order to investigate the electrochemical performance of the clip-coated separator, three different cathodes, infused sulfur cathode (ISC), vapor infused sulfur cathode (VISC), and ball-milled sulfur cathode (BMSC), have been used in this study. Figure 4a reports results from galvanostatic cycling studies of the materials in a 1M LiTFSI DOL/DME electrolyte and with ISC. The lithium metal anode used in the study was pretreated in the manner reported in reference[30], by soaking it in an electrolyte containing $LiNO_3$ for 24 hours followed by rigorous drying in an Ar environment. It is immediately apparent that they offer significant advantages. Specifically, in the control case with the pristine (uncoated) separator, the capacity dropped to 360 mAh g$^{-1}$ after 100 cycles at a current rate of 0.5C (838 mA g$^{-1}$). However, when the separator was coated using the aforementioned clip configuration, a capacity of ~1000 mAh g$^{-1}$ was obtained. The cells with the clip configuration also exhibit superior performance at high current rates. High initial capacities of 1290 mAh g$^{-1}$ and 1160 mAh g$^{-1}$ were obtained at 1C and 2C with high capacity retentions of 63% and 65%, respectively. Figure 4b reports the voltage profile at different cycle numbers for the clip configuration at 0.5C. Two discharge plateaus can be seen over many cycles: The first plateau at 2.35V corresponds to the reduction of the elemental sulfur to high order LiPS, whereas the second plateau at 2.05V indicates the high order LiPS reduction into low order LiPS. Notwithstanding the absence of LiNO3 in the electrolyte, there is no evidence of shuttling and a coulumbic efficiency as high as 99.8 % is observed. The voltage profiles for the uncoated separator control are provided in Figure S15a. Not only are the capacities found to be lower but also the telltale extended charge process associated with LiPS shuttling is observed. Supplementary Figure S15b reports the corresponding voltage profile of the clip configuration at different current rates. It is clear that the voltage plateaus for the discharge and charge processes does not change when the current is increased by 2 or 4 times.

To separate out the Li-S performance improvements derived from pretreating the Li metal anode with those obtained from the clip coated separator, Figure 4c reports performance of Li-S cells with the clip-coated separator, but in which pristine Li-metal anodes and additive-free electrolyte are used. Comparison of the results with those in Figure 4a indicates that comparable capacity is obtained when a pristine lithium metal anode is used. The comparison of the voltage profile of the cells utilizing the pristine separator and the clip coating separator is shown in Supplementary Figure S16. To facilitate comparisons with literature results, we also performed studies using a conventional Li-S electrolyte containing 0.05M $LiNO_3$ as the additive, and the results are shown in Figure 4d and corresponding voltage profiles are shown in Supplementary Figure S17. Figure 4e reports the rate capability of the cells with the clip coated separator are also significantly improved, illustrating that the capacity of the cells can recover after high rate cycles of 1C, 2C, and 3C for 10 cycles respectively. Cyclic voltammograms shown in Figure 4f further confirms the stability of the cells in the additive-free electrolyte with pristine Li anode. The discharge and charge peaks are observed to remain at the same position over many cycles, indicative of the stable and reversible electrochemical reaction of sulfur. Furthermore, Figure 4g and Figure 4h show electrochemical performance of the clip coated/pristine separators with VISC in 0.3M $LiNO_3$ and no $LiNO_3$ additive in the electrolyte. VISC has an areal sulfur loading of 3.5 mg $cm^{-2}$ and a content of 68w%. Figure 4g and 4h show effectiveness of the clip coated separator with VISC, and increasing capacity at first several cycles can be observed as reported in literatures for high loadings of sulfur with a upper current collector which confirms that the clip configuration plays a crucial role as the effective upper current collector for harvesting LiPS. Also, Supplementary Figure S18 shows the series of the voltage profiles of Li-S cells with the clip coated/pristine separators and VISC in the electrolyte with/without $LiNO_3$. More, the clip coated separators are also tested in a harsh environment - BMSC which is made via simple ball-milling sulfur powder with a carbon matrix and has a high sulfur loading of 3.5 mg $cm^{-2}$ and 70% (Figure 4i and Figure 4j). The clear improvements of the electrochemical performance of Li-S with the clip coated separators, the equilibrium capacity of ~700 mAh g$^{-1}$, are shown considering the mass loading of the clip coated separator, ~130 μg cm$^{-2}$, and the conditions of the cathode. Supplementary Figure S19 shows the series of the voltage profiles of Li-S cells with the clip coated/pristine separators and BMSC in the electrolyte with/without LiNO$_3$. Overall, the electrochemical performance of the clip coated separators are thoroughly examined using pretreated/pristine Li anode, various cathodes, and the electrolyte with/without LiNO$_3$, and good electrochemical performance of the Li-S cells with the clip coated separators are achieved. Longer cycling performance of clip coated separators with ISC and VISC without LiNO$_3$ in the electrolyte is shown in Figure 4k, and stable performance is observed >200 cycles at 0.5 C, which is remarkable with such high sulfur loading and LiNO$_3$ free electrolyte.

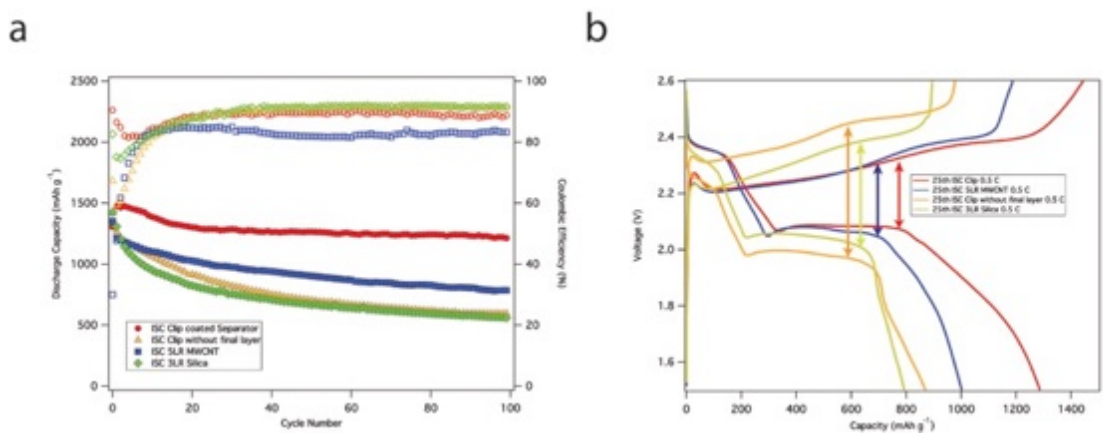

**Figure 5 | Electrochemical performance of the clip multilayer parts. a**, Cycling performance of the pristine Li anode with the clip coated separator, clip coated separator without final MWCNT layer (Same structure as shown in Figure 3d), five layers of MWCNT coated separator, and three monolayers of silica nanospheres coated separator Li-S cell with ISC at 0.5 C. **b**, Discharge-charge voltage profiles of pristine Li anode with clip coated separator, clip coated separator without final MWCNT layer, five layers of MWCNT coated separator, and three monolayers of silica nanospheres coated separator Li-S cell with ISC for 25$^{th}$ cycle at 0.5 C.

In order to investigate the effect of each compartment in the clip configuration, we compared the electrochemical performance of the clip with five layers of MWCNT coating, three layers of silica, and a clip configuration without the final electrical path MWCNT layer, which is the equivalent structure shown in Figure 3d (Figure 5a). Significantly, it is noted that without the final MWCNT coating to complete the clip, similar electrochemical performances are observed in Li-S cells using the multifunctional MWCNT-SiO$_2$ coatings, compared to those based on separators coated with three monolayers of silica. These results underscore the importance of the clip configuration in complementing LiPS adsorption achieved with $SiO_2$ coatings, with utilization of the trapped LiPS made possible by the MWCNT coating layers. They also validate our hypothesis that a good electrical conductive path is required to efficiently entrap and utilize dissolved LiPS. Figure 5b compares the voltage profiles of the cell with the clip configuration done in different steps. Consistent with the previous observation, the capacity is seen to increase progressively as the clip components are sequentially added to complete the structure: 561 mAh $g^{-1}$ at $100^{th}$ cycle, 596 mAh $g^{-1}$ at $100^{th}$ cycle, 785 mAh $g^{-1}$ at $100^{th}$ cycle, and 1214 mAh $g^{-1}$ at $100^{th}$ cycle when 3LR silica, the multifunctional MWCNT-$SiO_2$, 5LR MWCNT, and clip are coated on the separator, respectively. Another important observation is that the overpotential of the cell substantially declines when the final layer is involved, which confirms our hypothesis that the silica surface traps LiPS in the separator, which overtime reduces the electrolyte conductivity. Reutilization of the LiPS in the clip configuration eliminates this problem and reduces the overpotential correspondingly.

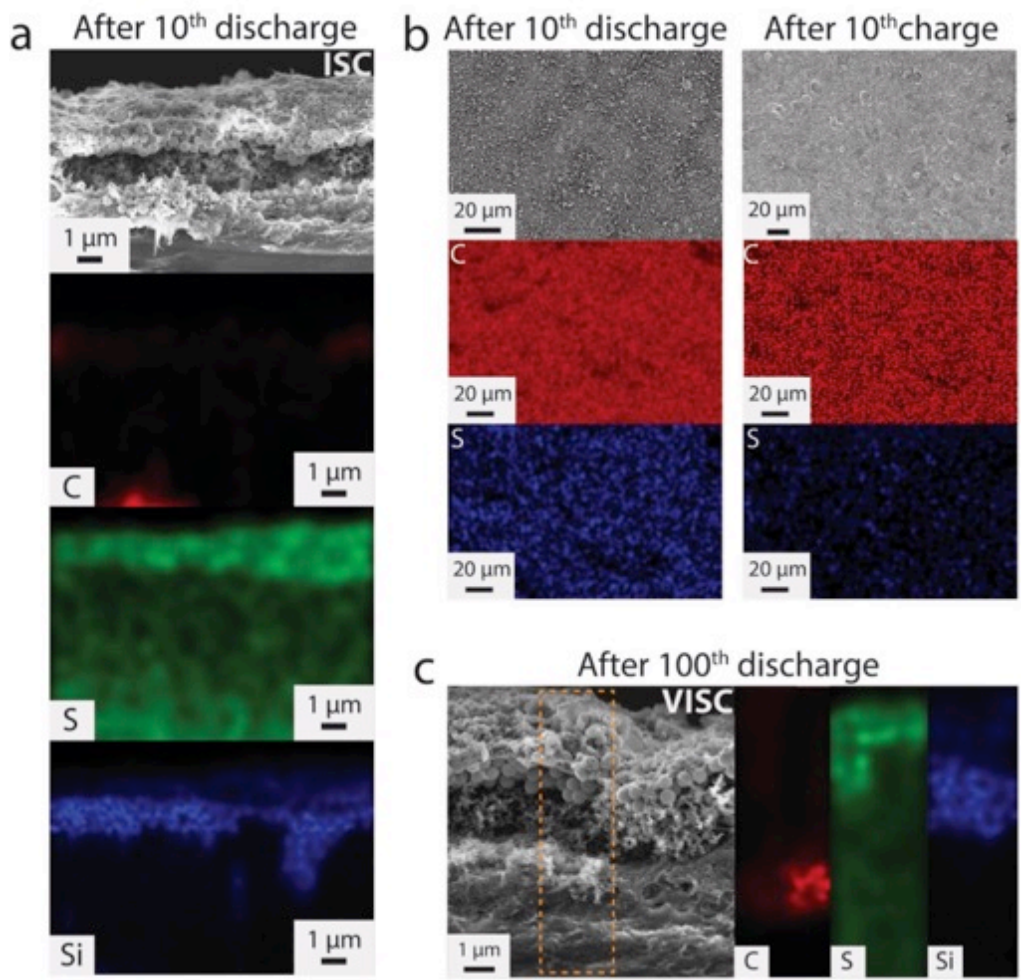

**Figure 6 | Clip coated separator morphology and elemental mappings after several cycles. a**, Cross sectional SEM image of the clip coated separator with ISC after $10^{th}$ discharge with carbon, sulfur, and silicon maps. **b**, Top view SEM image of the clip coated separator with ISC after $10^{th}$ discharge/charge with carbon and sulfur maps. **c**, Cross sectional SEM image of the clip coated separator with VISC after $100^{th}$ discharge with carbon, sulfur, and silicon maps.

Finally, we investigated the morphology of the coating surface on the clip-coated separator after cycling using SEM. Figure 6a shows the morphology of the cross sectional separator after $10^{th}$ discharge and corresponding energy-dispersed X-ray spectroscopy (EDXS) measurements of elemental mapping of carbon, sulfur, and silicon, in which silicon layer and sulfur distribution in the clip structure is clearly seen. The low x-ray intensity for carbon is due to high coverage of sulfur after the discharge, indicating carbon layers are efficiently trapping dissolved LiPS. The structure of the clip coating is preserved upon discharge and charge, indicating the robust properties of the coating in both mechanical and chemical aspects. Also EDXS measurements of top view of the clip coated separator show a uniform distribution of carbon and sulfur elements (Figure 6b). The fact that the amount of sulfur decreased during the charge process further substantiates the ability of the coating to reutilize the adsorbed species. In addition, the morphology of the clip coated separator with VISC after $100^{th}$ cycles is shown in Figure 6c. This double confirms that the multifunctional coatings remain robust after large number of cycles with the presence of high loading and content of sulfur in the Li-S cells. We have also observed decrease in internal resistance of the Li-S cell for clip coated separator compared to pristine separator (See Supplementary Figure S20). The decreased impedance indicated that the clip design is able to facilitate the electron transfer even the insulating silica particles are involved.

**Conclusion**

In summary, we have demonstrated two new versatile coating methods, LBSDC and LBS, for creating surface films that utilize the surface tension gradient to create well-ordered monolayer films at an air/water interface. The methods allow multifunctional coatings to be created in a range of designs using a wide selection of individual materials, as well as material combinations, on a variety of substrates, without the need for chemical binders. The utility of the approach is illustrated using the polypropylene membrane separator, Celgard, commonly employed in Li-S batteries as a substrate. Through systematic studies, it is shown how mono-functional coatings based on different metal oxides and carbon influence reutilization of dissoleved lithium polysulfide species. An unusual coating configuration termed the "clip", created by stacking an incomplete, but well-formed layer of $SiO_2$ particles between two complete layers of carbon is used to illustrate both the versatility of the method to create multifunctional coatings with good spatial control and the effectiveness of such coatings in battery separators. In particular, the *clip* coated separator is observed to exhibit largely improved active material utilization, enhanced capacity retention over extended charge/discharge cycling, and attractive high rate capability. These observations are explained in terms of the ability of the multifunctional coatings to simultaneously adsorb and trap LiPS created at the cathode, without loosing electrochemical access to the materials. The new coating approach and configurational design of coating materials synergistically work together to advance Li-S cells and allow us to investigate and optimize the different coating structures.

**Methods**

*Synthesis*

**Silica nanospheres**: Silica nanospheres were synthesized by the Stober synthesis technique[49]. In this method, 10 ml of amonium hydroxide, 10 ml of water and 75 ml of ethanol are taken in a round bottom flask and stirred using a stir bar to ensure proper mixing. Under smooth stirring conditions, 5.6 ml of tetraethyl orthosilicate (TEOS) was added drop-wise. After 12 hours of stirring, the prepared monodispersed silica nanospheres was purified by alternate centrifuging and sonication in an ethanol-water mixture until the colloidal solution reaches a stable pH of 7. The size of the prepared silica was characterized using the scanning electron microscopy (SEM) technique. The resulting silica nanosphere was determined by means of dynamic light scattering (DLS) and SEM analysis to be approximately 350 nm in diameter.

**Suspensions**: A Langmuir-Blodgett (LB) film forming suspension was comprised of 1 wt% of a nanomaterial in pure ethanol (Decon, 200 Proof). Each 1wt% of silica nanosphere, titania nanopowder (Rutile, 99% purity, Advanced Materials™), multi-walled carbon nanotube (L 6-9 nm x 5 μm, >95% carbon, Sigma Aldrich), Ketjen-Black carbon (Akzo Nobel), and Super-P carbon (TIMCAL) was dispersed in pure ethanol. Then each of the suspension was sonicated for 30 minutes to enhance the dispersion of the particles. Note that the film quality is not sensitive to the weight composition of the nanomaterials in ethanol solvent; 0.5 to 3 wt% of the nanomaterial suspensions yielded the same quality films. However, the film quality heavily depends on the distribution of the dispersed particle sizes; big aggregates or clusters will form a defect during the self assembled LB film.

**Surfactant**: The SDS surfactant is made by dissolving 3 wt% of sodium dodecyl sulfate (SDS) in DI water.

**Sulfur infused KB composite**: Sulfur infused in Ketjen Black carbon composite was prepared by an infusion method. First, sulfur powder and Ketjen Black carbon (2.2:1 by weight) were placed in a hollow glass vial in Ar atmosphere. Then, the end of the glass vial was sealed to avoid water moisture during the infusion process. The composite contained glass vial was heated to 155 °C for 12 hours to infuse active sulfur into the pores of Ketjen Black carbon and subsequently cooled to room temperature. The resulting composite had a sulfur content of 66 wt% (Supplementary Figure S21).

**Cathode preparation**: ISC - The sulfur infused Ketjen Black composite (77 wt%) was mixed with Super P (8 wt%) and 10 wt% polyvinylidene fluoride (Sigma Aldrich) dissolved in N-methyl-2-prolidone (15 wt%) in N-methyl-2-prolidone (Sigma Aldrich), and the mixture is ball-milled at 50 rev s$^{-1}$ for 30 minutes. The resulting viscous slurry was coated onto a carbon sprayed aluminum foil as a current collector using doctor-blade method. The coated slurry is then dried in a convection oven at 60 °C for 5 hours. The prepared electrode is cut into a circular disk, and the electrode has sulfur loading of 0.5 – 0.55 mg cm$^{-2}$ with 50 wt% of active sulfur per cathode. After including conductive carbon components in the clip coated separator (excluding the mass of silica nanospheres), the active sulfur content remains at 47.5%. BMSC - The 70 wt% sulfur powder (Sigma Aldrich) was mixed with Super P (20 wt%) and 10 wt% polyvinylidene fluoride (Sigma Aldrich) dissolved in N-methyl-2-prolidone (15 wt%) in N-methyl-2-prolidone (Sigma Aldrich), and the mixture is ball-milled at 50 rev s$^{-1}$ for 30 minutes. The resulting viscous slurry was coated onto a carbon sprayed aluminum foil as a current collector using doctor-blade method. The coated slurry is then dried in a convection oven at 60 °C for 5 hours. The prepared electrode is cut into a circular disk, and the electrode has sulfur loading of 3.5 mg cm$^{-2}$ with 70 wt% of active sulfur per cathode. VISC – this is a carbon-sulfur cathode created by infusion of sulfur in the vapor phase into a carbon fiber matrix.[21] Sulfur cathodes are prepared by coating the composite material onto a carbon coated Al foil. The cathode has a high sulfur loading (68w%) and a high areal density of sulfur (3.5 mg cm$^{-2}$). After including conductive carbon components in the clip-coated separator (excluding the mass of silica nanospheres), the sulfur contents for BMSC and VISC are 69.31% and 67.35%, respectively. All measurements reported in the study utilize a cathode size of 1.26 cm$^2$.

*Coating process*
**Separator**: The commercial polypropylene separator (Celgard 2500) was cut into a 1.6 cm diameter circular disk. The separator was placed onto a 1.8 x 1.8 cm microscope cover glass and the ends taped with Kapton tape to control the coating location (Supplementary Figure S22).

**Single component separator coating**: Mono/multi layers of Silica nanospheres, multi-walled cabon nanotubes, Ketjen Black carbon, and Super P carbon were coated on the separator using LBS method. Prepared separators are washed with tap water to flush out any impurities stuck onto the surface. One drop of isopropanol (IPA) is applied onto the separator or separator with coating layers to uniformly wet the surface of the separator with water, and the excess IPA on the separator is diluted with water. Then, the fully wetted separator is immersed in water. One of the suspensions is then injected at the surface of the water until more than half of the surface is saturated with the desired nanomaterial; the separator is subsequently raised up to transfer the film followed by a constant injection of the suspension. After that, the coated separator is dried on a hot plate at 110 °C for 30 seconds. Note that no IPA wetting step is required for the silica nanosphere coatings. The single layer coating process is repeated until the desired number of layers is achieved. After the final layer coating, the whole separator is dried on the hot plate for one minute at 110 °C.

**Clip coating**: The clip configuration coating is comprised of five coating layers of MWCNT, one coating layer of SP, three monolayers of silica nanospheres, and one final coating layer of MWCNT. The first five coating layers of MWCNT are coated in the same manner as the single component separator coating. Then, one layer of SP, which acts as an adhesion layer for the silica nanospheres, is coated on top of MWCNT using the LBS method. 80% of the MWCNT and SP carbon coated separator is covered with three monolayers of silica nanospheres using the LBSDC method. During the silica coating, no IPA is used after each coating. For the final layer, the remaining 20% of the separator is wetted with IPA. And after the dilution of IPA, the whole surface of the separator is coated with one coatinglayer of MWCNT. The clip coated separator is then dried on the hot plate at 110 °C for one minute.

*Battery preparation*
**Li anode pretreatment**: Li metal foil was cut into a 1.27 cm diameter circular disk, and the Li metal disks are completely soaked in 0.5M $LiNO_3$ (Sigma-Aldrich) 1,2-dimethoxyethane (DME, Sigma Aldrich) and 1,3-dioxolane (DOL, Sigma Aldrich) (1:1 v/v) electrolyte solutions for 24 hours [30]. Then, the pretreated Li metals were rigorously dried in air/oxygen-free Ar environment.

**Electrolyte**: Three electrolytes are prepared for this study: **i)** 1M Bis(trifluoromethane)sulfonamide lithium salt (LiTFSI, Sigma Aldrich) in DME:DOL (1:1 v/v) electrolyte, **ii)** 1M LiTFSI with 0.05M LiNO$_3$ in DME:DOL (1:1 v/v), and **iii)** 1M LiTFSI with 0.3M LiNO$_3$ in DME:DOL (1:1 v/v).

**Cell assembly**: CR2032-type Li-S coin cells are assembled with the polypropylene separators with coating layers, Li-foil disks, as prepared Li-S cathode, stainless-steel springs and spacers, and the electrolytes. A total of 40 µL of the electrolyte is used per cell. The first 20 µL of the electrolyte is added to the coating layers of the separator. Then, the cathode is placed onto the electrolyte-wetted separator facing the coating layers. Another 20 µL of the electrolyte is applied to the other side of the separator and pretreated or pristine Li metal disk is placed. Then the spacer and spring are used to sandwich the anode/separator/cathode, and pressure of 15 MPa is applied to punch the cell. The assembled cell is rested for about 15 minutes before testing. Cell assembly was carried out in an Ar filled glove-box (MBraun Labmaster). The room-temperature cycling characteristics of the cells were evaluated under galvanostatic conditions using Neware CT – 3008 battery testers and the electrochemical process in the cells were studied by cyclic voltammetry using a CHI600D potentiostat.

*Characterization*

**Langmuir-Blodgett trough:** Surface pressures of 350 nm Silica colloids, MWCNT, KB, SP, and SDS surfactant are measured using conventional LB trough (KSV NIMA L & LB Troughs). The trough has dimension of 7.5 cm x 32.4 cm (Supplementary Figure S2). The trough was cleaned using pure ethanol and DI water and fully dried with nitrogen gas. The trough was filled with DI water and a 0.5 ml of suspension is injected at the ends of the trough to float particles. After the injection of the suspension, rest time of ~7 minutes was needed to evaporate excess ethanol from the suspension. Then the resulting film is compressed at the rate of 3 mm min$^{-1}$ to collect the surface pressure profiles. For obtaining the pressure profile of LBDSC and LBS, ~35 cm$^2$ and ~25 cm$^2$ areas are set to mimic actual coating process occurs at the surface from the 50 ml glass beaker. To collect LBDSC surface profile, ~35 cm$^2$ of the surface is saturated with the silica colloids and rested about 5 minutes to evaporate remaining ethanol. Then, 5 µL of the surfactant is added and the pressure profile is collected at the compression rate of 3 mm min$^{-1}$. For LBS surface profiles, ~25 cm$^2$ area is fully covered by MWCNT, KB, and SP, and without the rest time, the surface pressure profiles are collected at the compression rate of 3 mm min$^{-1}$. The surfactant surface pressure profile is measured at four different areas (7 cm$^2$, 11 cm$^2$, 19 cm$^2$, 38 cm$^2$) without compressing the barriers, and the 5 μL of the surfactant surface pressures are measured over time.

**Galvanostatic charge/discharge**: Neware battery testing system is used to perform cycling testing of the Li-S cells. 1.5V to 2.6V and 1.7V to 2.6V voltage windows are used for without/with LiNO$_3$ electrolyte systems, respectively. 1.7V to 2.6V voltage window is chosen for the LiNO$_3$ system to preserve LiNO$_3$ passivation layer on the Li.

**CV**: CHI600D potentiostat is used to perform cyclic voltammetry analysis of the cell. 0.1 mV s$^{-1}$ scan rate with the voltage window of 1.5V to 2.6V are chosen as the parameters.

**SEM**: The morphology of the coating layers on the separator is analyzed using Keck scanning electron microscope (LEO 1550 FESEM) at 3 kV acceleration voltage.

**EDXS**: Energy dispersive X ray spectroscopy (EDXS) is performed on Keck scanning electron microscope to investigate the chemistry of the coatings on the separator.

**TGA**: Thermogravimetric analysis (TGA) was used to determine the content of sulfur in the S-KB composite. Morphologies of the electrodes were studied using Keck SEM.

**ACI**: Alternating current impedance (ACI) was measured versus frequency using a Novocontrol N40 broadband dielectric spectroscopy.

**Acknowledgements**

The authors acknowledge support of the National Science Foundation Partnerships for Innovation Program (Grant No. IIP-1237622). Electron microscopy, X-ray diffractometry, X-ray spectroscopy facilities and optical spectrometers available through the Cornell Center for Materials Research (CCMR) were used for this work (NSF Grant DMR-1120296). MK acknowledges partial support from the Korea Institute of Science and Technology (KIST) Institutional Program.

**Additional information**

Supplementary information. Reprints and permissions information. Correspondence and request for materials should be address to L.A.A.

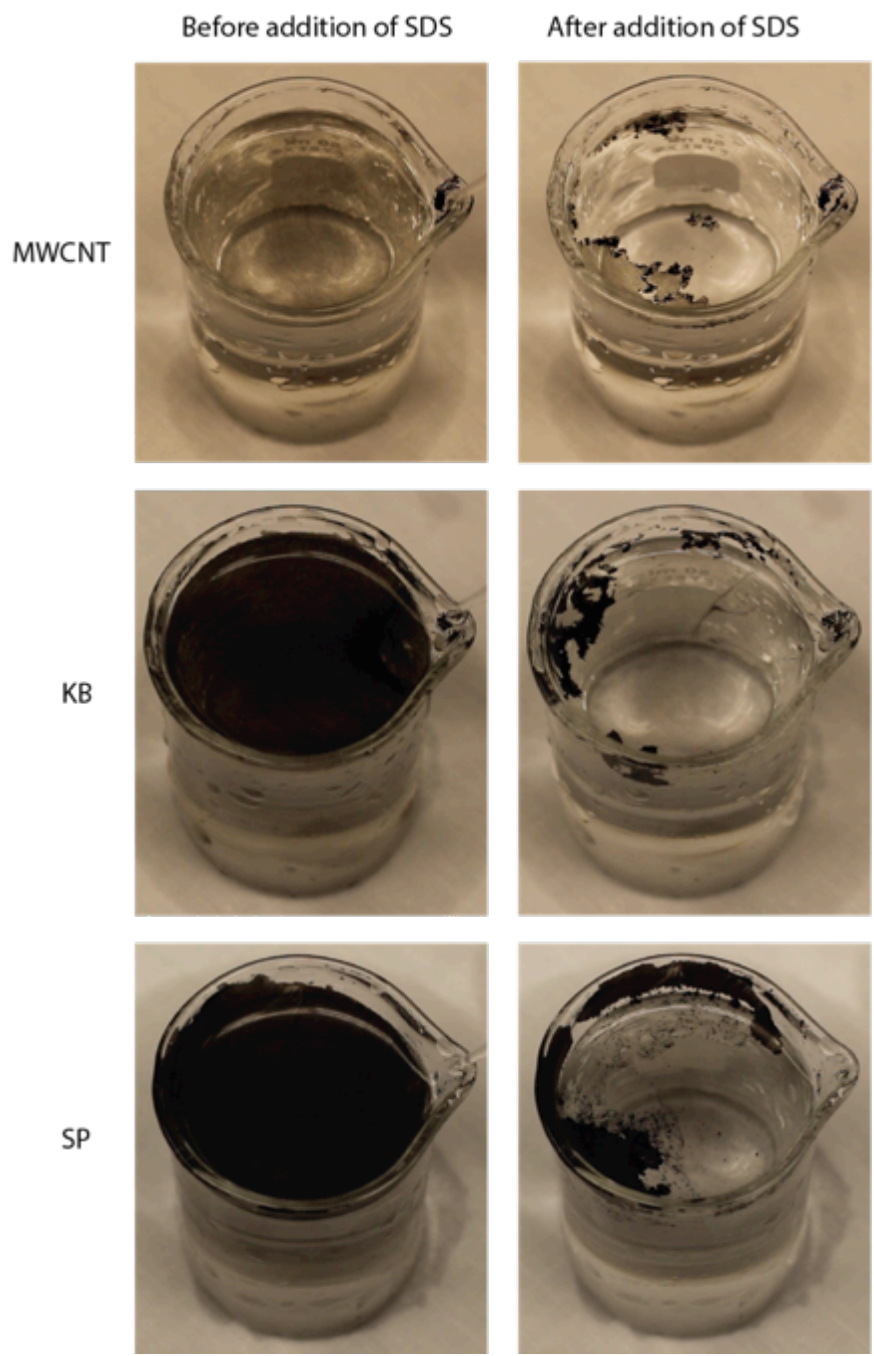

**Supplementary Figure S1: Collapsing self-assembled LB film in the presence of SDS surfactant.** The self-assembled MWCNT, KB, and SP films are formed at the surface of water using LBS coating method, and images of the films before and after adding a one-drop of SDS surfactant at the neck of the beaker are shown.

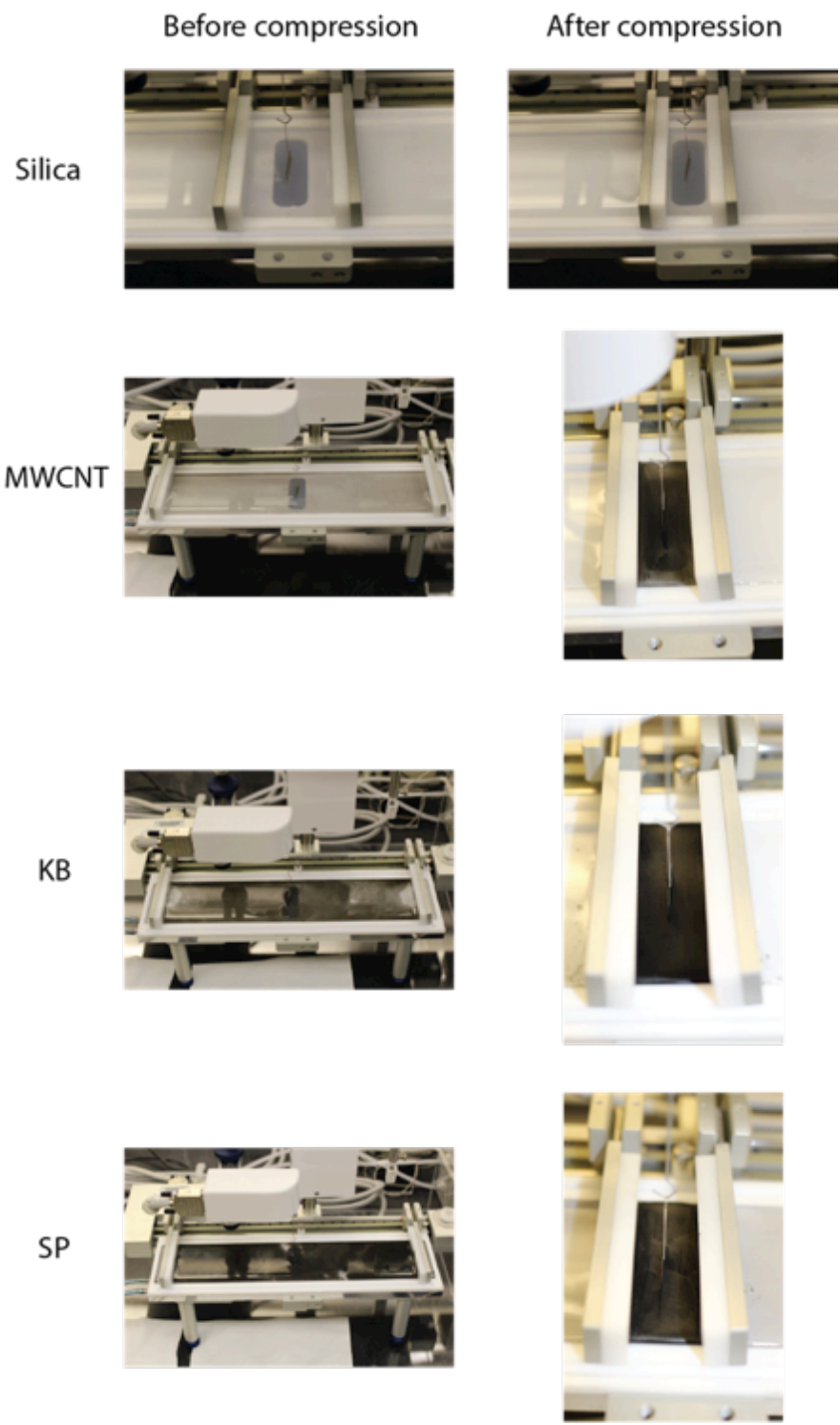

**Supplementary Figure S2: Langmuir-Blodgett trough experiment for surface pressure measurements.** Images of compressing silica nanospheres, MWCNT, KB, and SP are shown. Note that no materials are lost during the compression of the LB films, and folding of the films is observed instead of particles sinking.

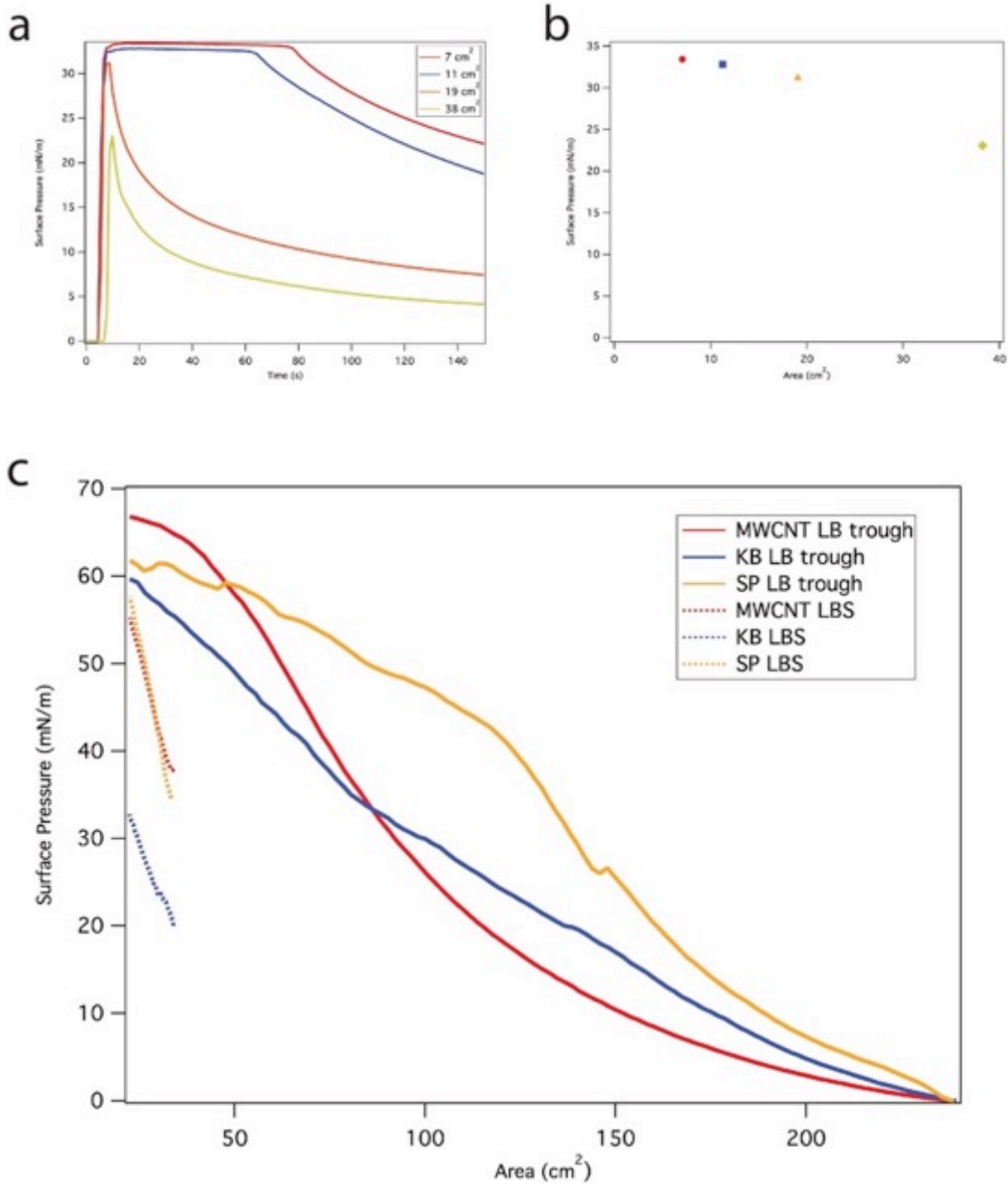

**Supplementary Figure S3: SDS surfactant and carbon surface pressure profiles.** (a) Surface pressures of SDS surfactant versus time for the surface area of 7 cm$^2$, 11 cm$^2$, 19 cm$^2$, and 38 cm$^2$. (b) Maximum surface pressures exerted by the surfactant for the surface area of 7 cm$^2$, 11 cm$^2$, 19 cm$^2$, and 38 cm$^2$. (c) MWCNT, KB, and SP surface pressure profiles obtained from conventional LB trough and LBS methods.

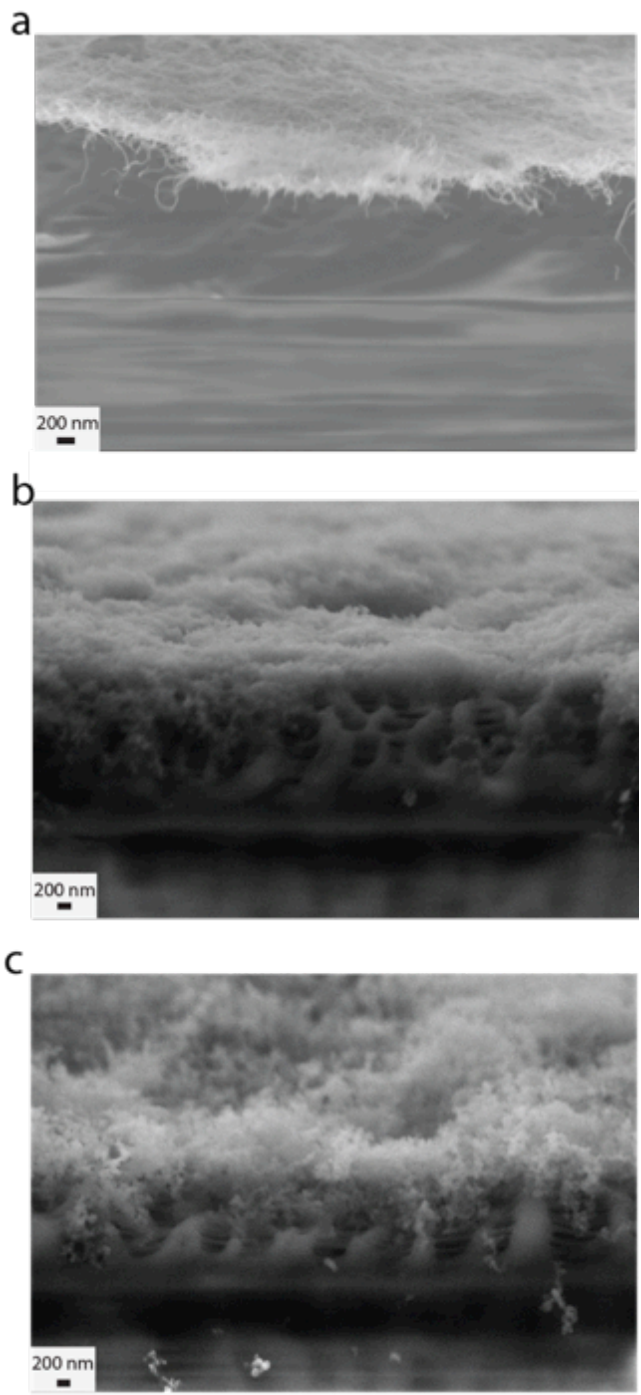

**Supplementary Figure S4: Thickness of single coating layer carbon films.** Cross-sectional SEM images of a single layer coating of (a) MWCNT, (b) KB, and (c) SP on the separator using LBS method.

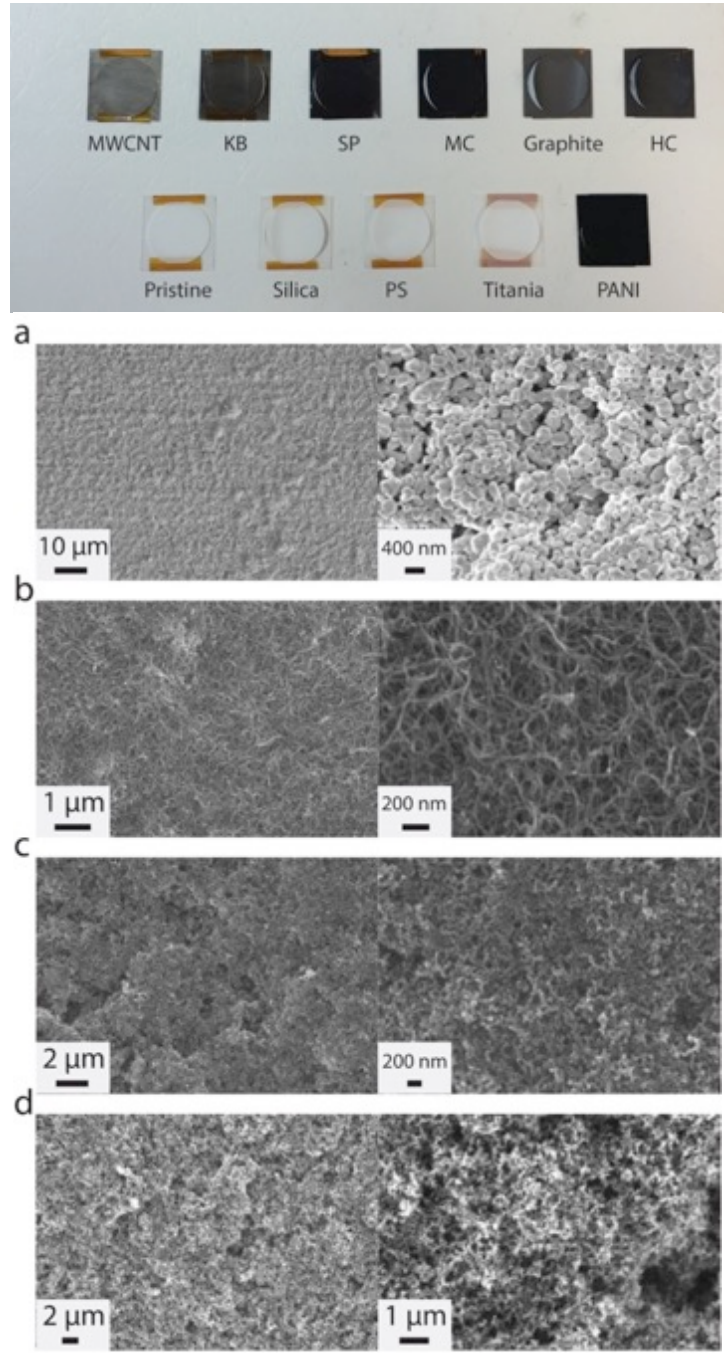

**Supplementary Figure S5: Demonstration of coating various materials onto the separator using LBS method**. Images of single coating layer of Multi-walled carbon nanotube (MWCNT), Ketjen Black carbon (KB), Super P carbon (SP), Microporous carbon (MC), Graphite, Hard carbon (HC), 350 nm silica nanospheres, 1μm polystyrene spheres (PS), titania nanopowder, and polyaniline (PANI) on the separator are shown. SEM images of three coating layers of (a) titania nanopowder, (b) MWCNT, (c) KB, and (d) SP coated on the separator.

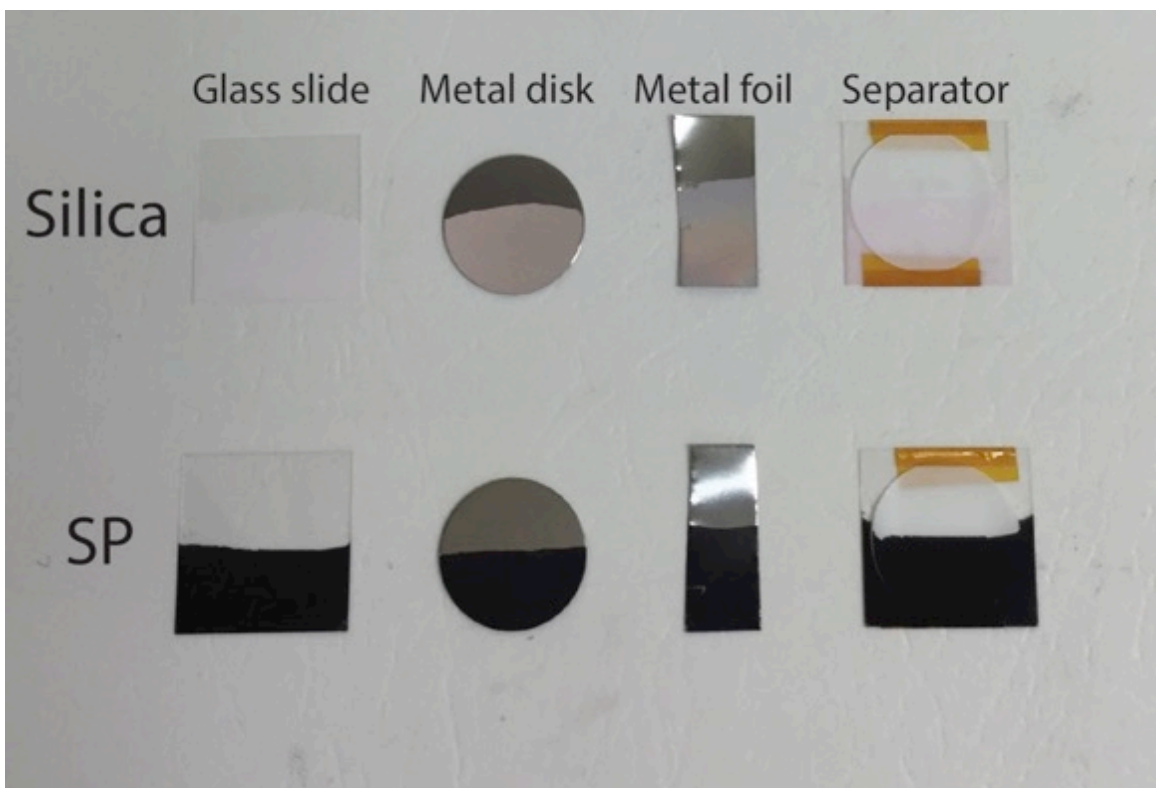

**Supplementary Figure S6: Uniform substrate coating demonstration.** Images of fluorescent 1μm silica nanosphere and Super P carbon coatings on various uniform substrates using LBS coating method are shown. Note that other materials can also be coated on these substrates.

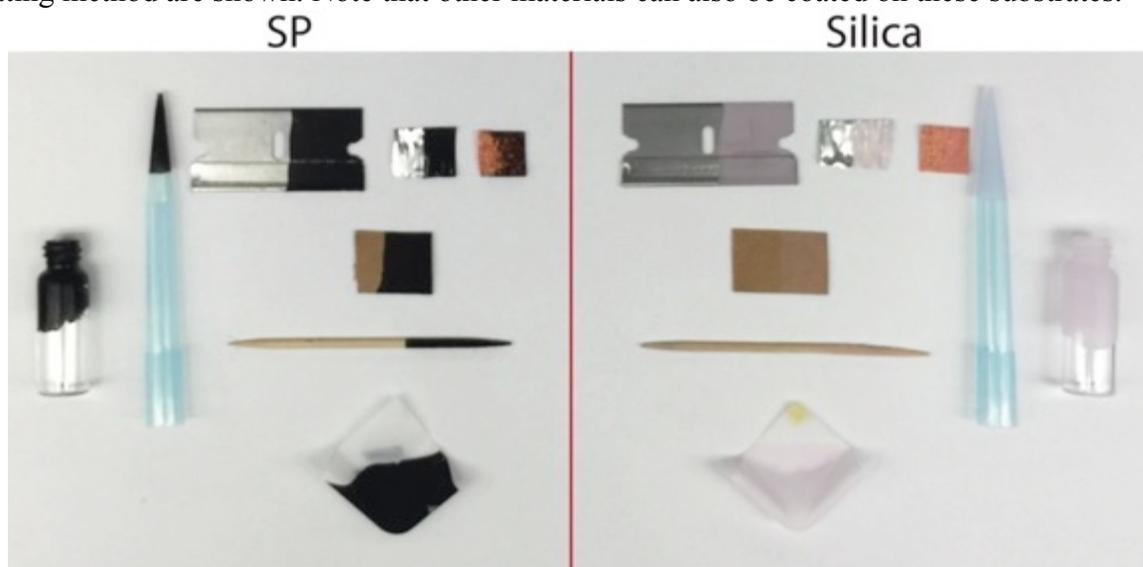

**Supplementary Figure S7: Non-uniform substrate coating demonstration.** Images of Super P carbon and fluorescent 1μm silica nanosphere coatings on razor blade, rough metallic strip, metallic foam, paper, wood toothpicks, curved plastic, tip of glass vials, and micropipette tips using LBS coating method are shown. Note that other materials can also be coated on these substrates.
36

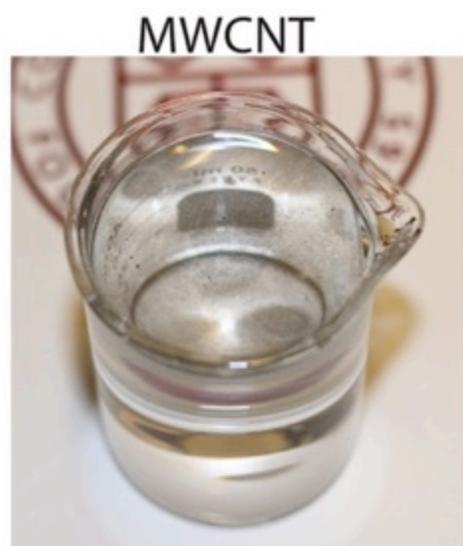
MWCNT
~5 μg cm$^{-2}$

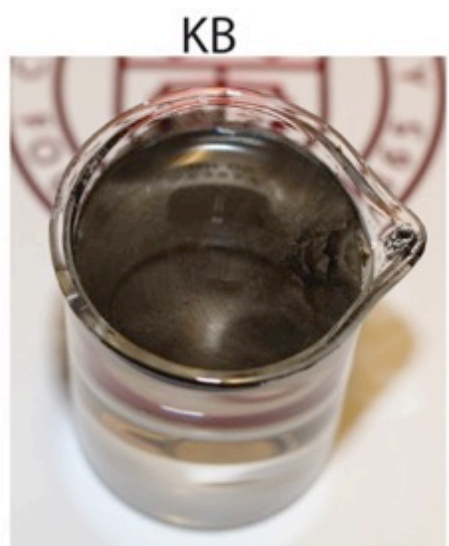
KB
~17 μg cm$^{-2}$

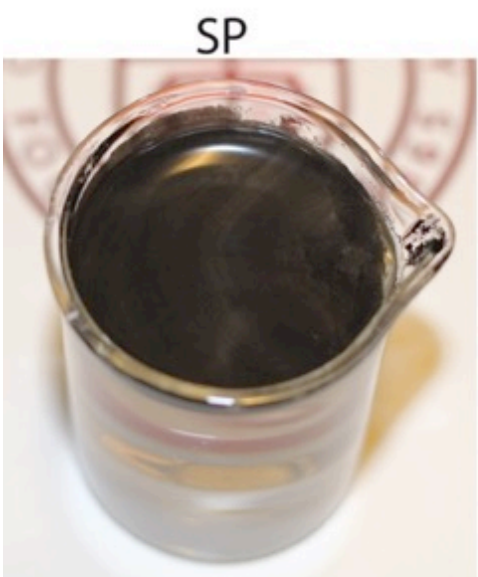
SP
~20 μg cm$^{-2}$

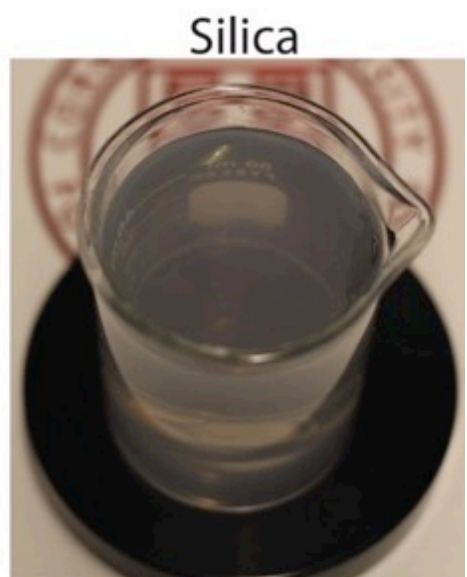
Silica
~25 μg cm$^{-2}$

**Supplementary Figure S8: Self-assembled Langmuir-Blodgett films**. Single coating layer of multi-walled carbon nanotube (MWCNT), Ketjen Black carbon (KB), Super P carbon (SP), and silica nanosphere is shown with gravimetric areal density of the LB films.

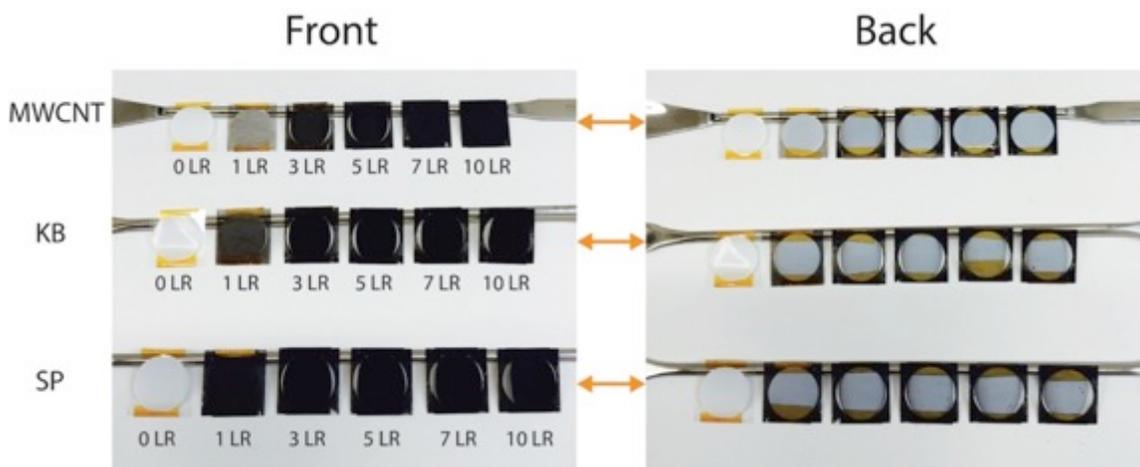

**Supplementary Figure S9: Single-sided separator coating demonstration.** Images of zero to ten coating layers of multi-walled carbon nanotube (MWCNT), Ketjen Black carbon (KB), and Super P carbon (SP) on the separator using LBS method are shown.

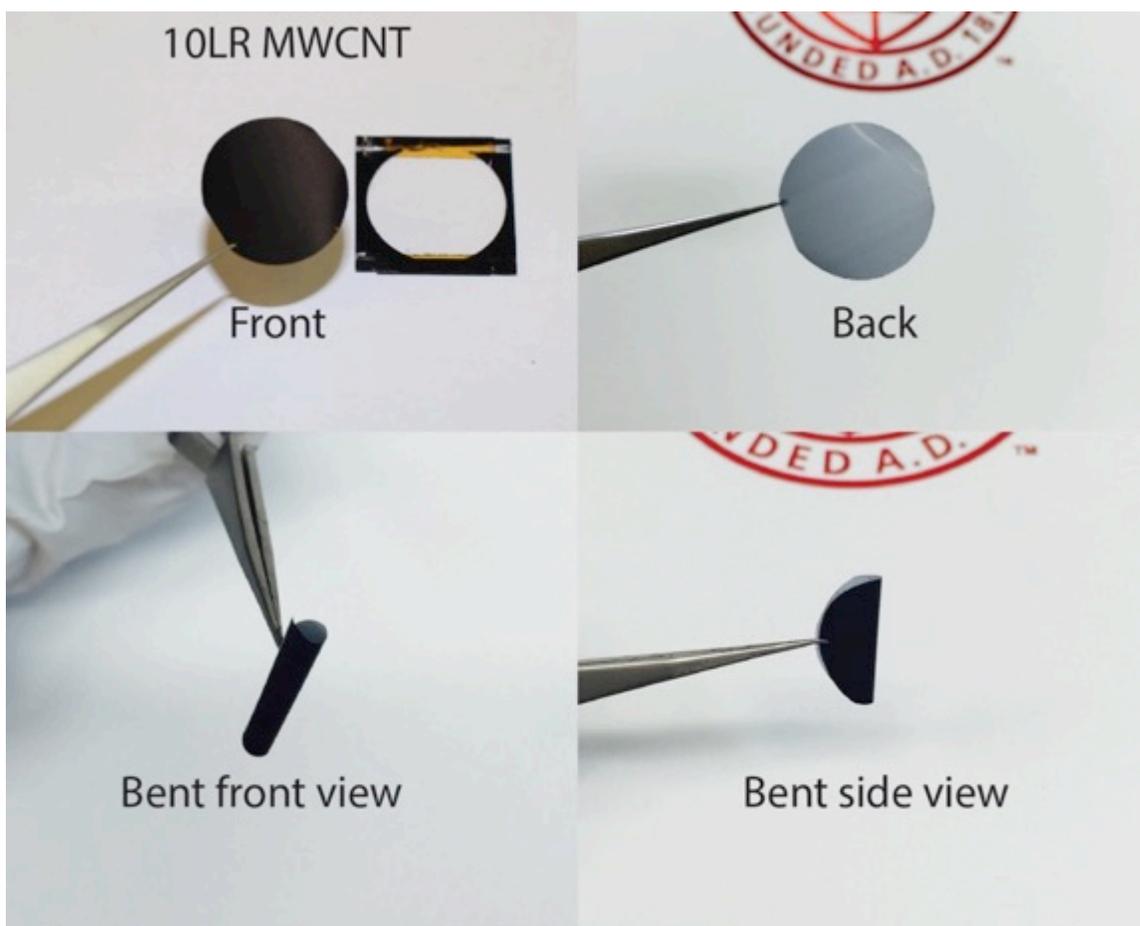

**Supplementary Figure S10: MWCNT separator coating quality demonstration.** Ten coating layers of MWCNT on the separator are shown.

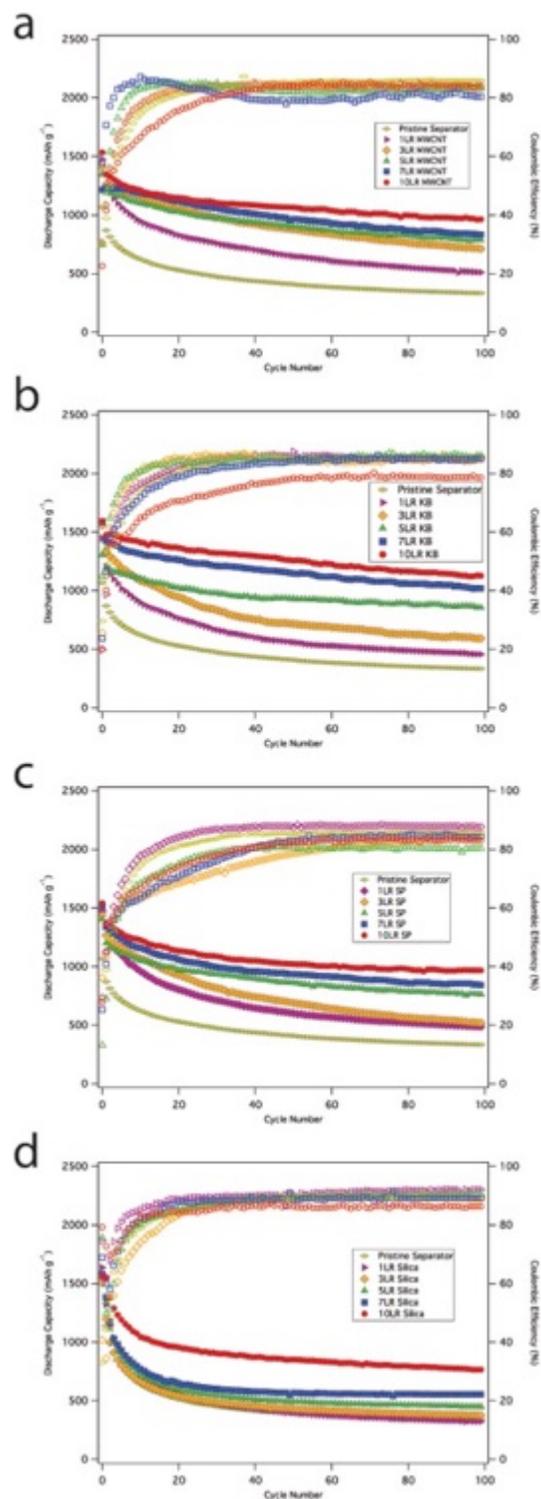

**Supplementary Figure S11: Electrochemical performances of silica and carbon coated separators with ISC**. Electrochemical performance of zero to ten coating layers of (a) MWCNT, (b) KB, (c) SP coated separators Li-S cells at 0.5 C. (d) Cycling performance of zero to ten monolayers of silica nanospheres coated separators Li-S cells at 0.2 C.

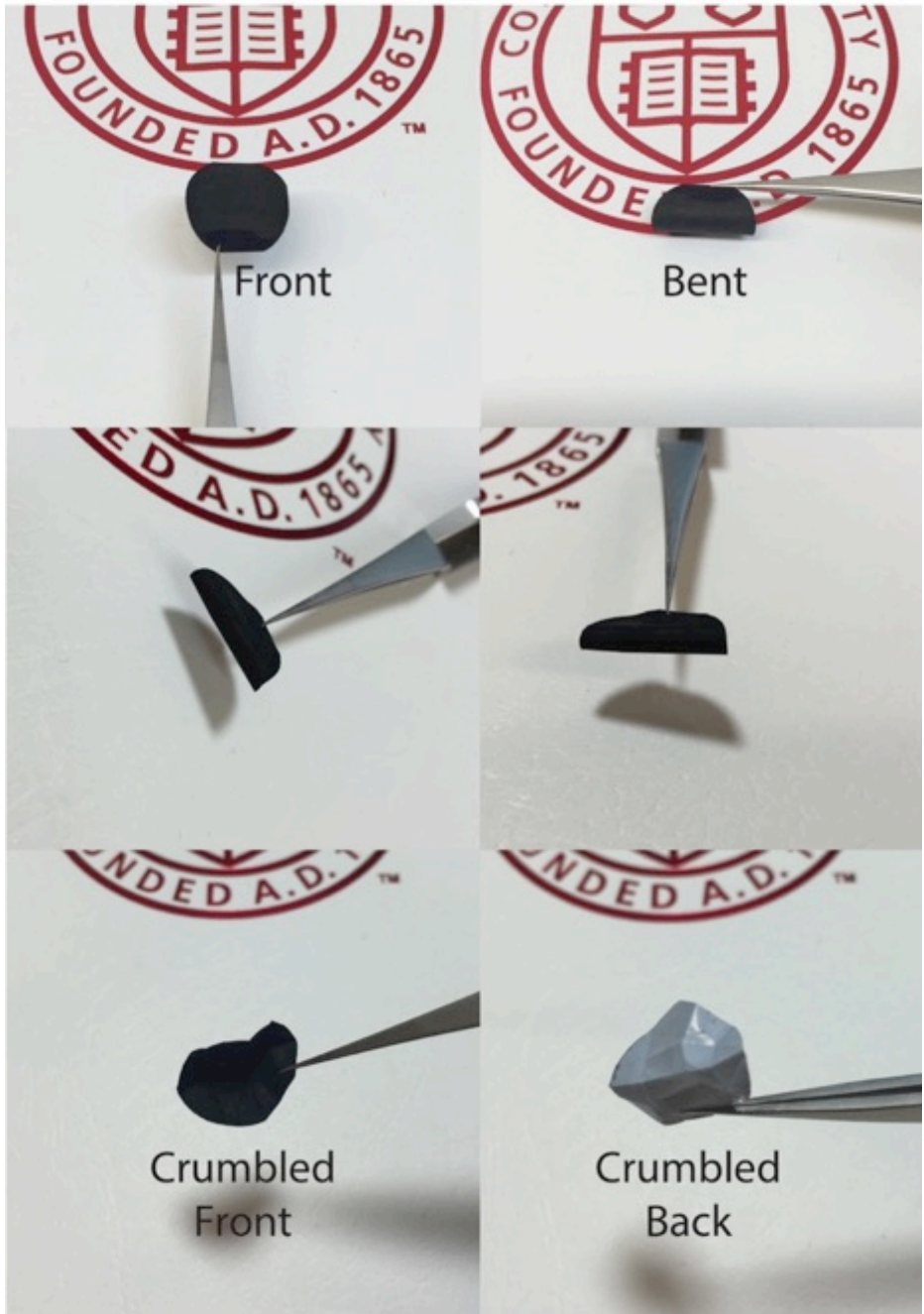

**Supplementary Figure S12: Coating quality and mechanical strength demonstration of clip coated separator.**

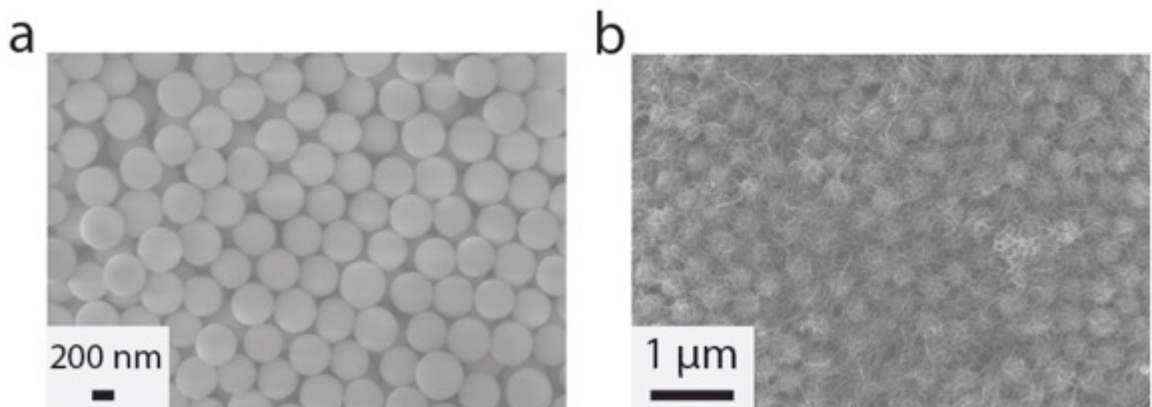

**Supplementary Figure S13:** Top-view SEM images of (a) three monolayers of silica nanospheres coated on top of SP-MWCNT coated separator and (b) the clip.

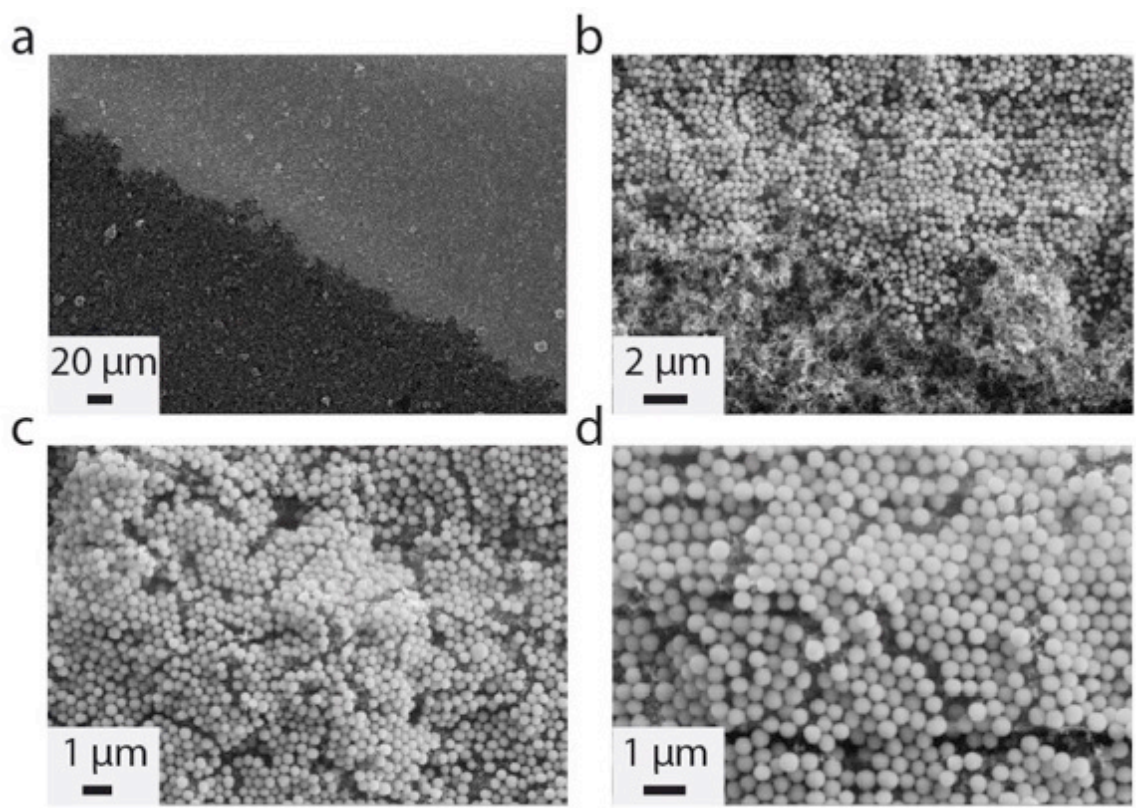

**Supplementary Figure S14: SEM images of a monolayer of silica nanospheres coated on SP-MWCNT layer.** Silica-SP layer boundary region (a) wide-view and (b) close-view. A monolayer of silica coating on top of the SP-MWCNT layer (c, d).

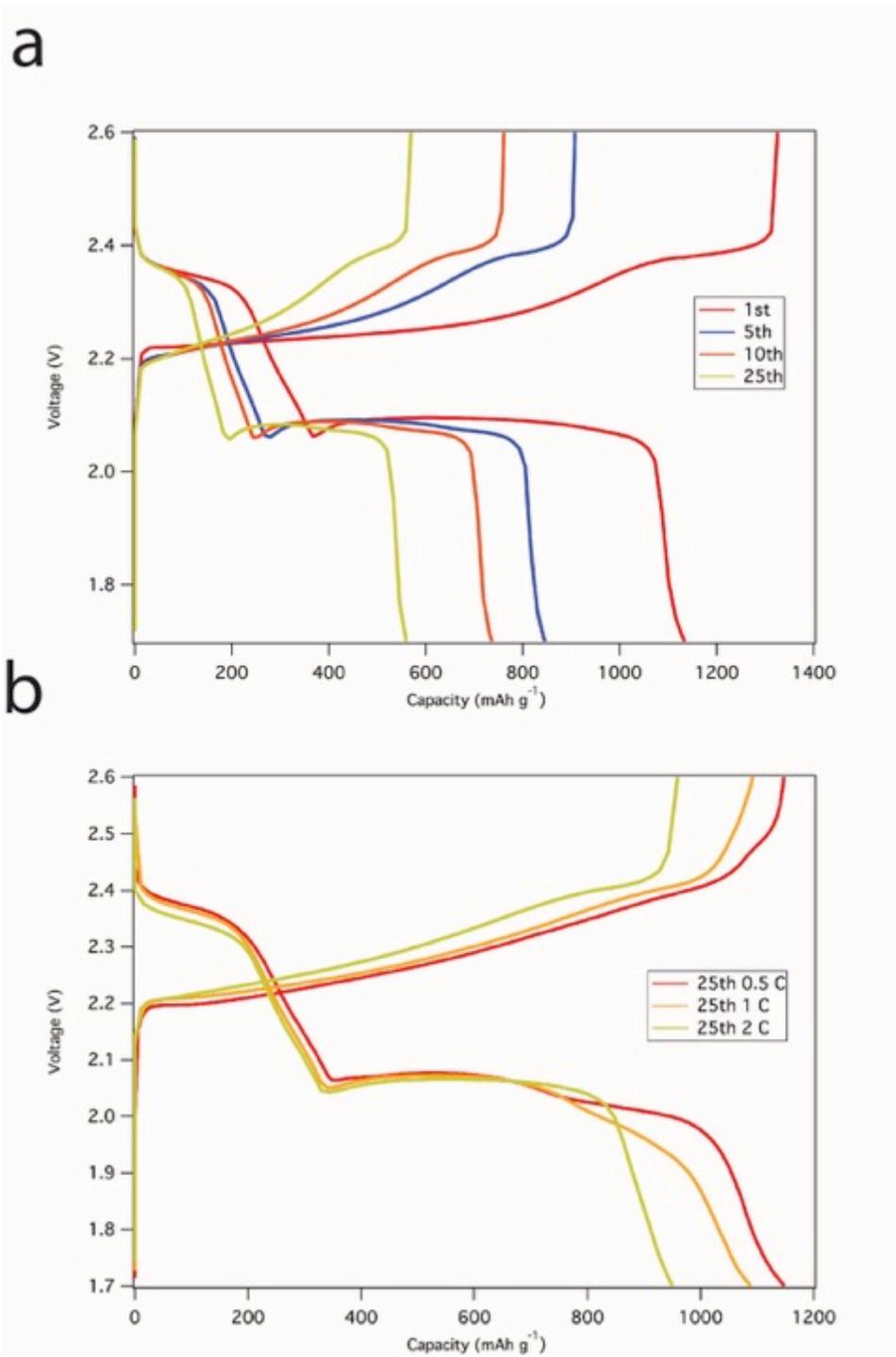

**Supplementary Figure S15:** Discharge-charge voltage profiles of clip coated separator Li-S cell with the pretreated Li anode and with ISC for (a) various cycles at 0.5C and (b) various C rates at 25$^{th}$ cycle.

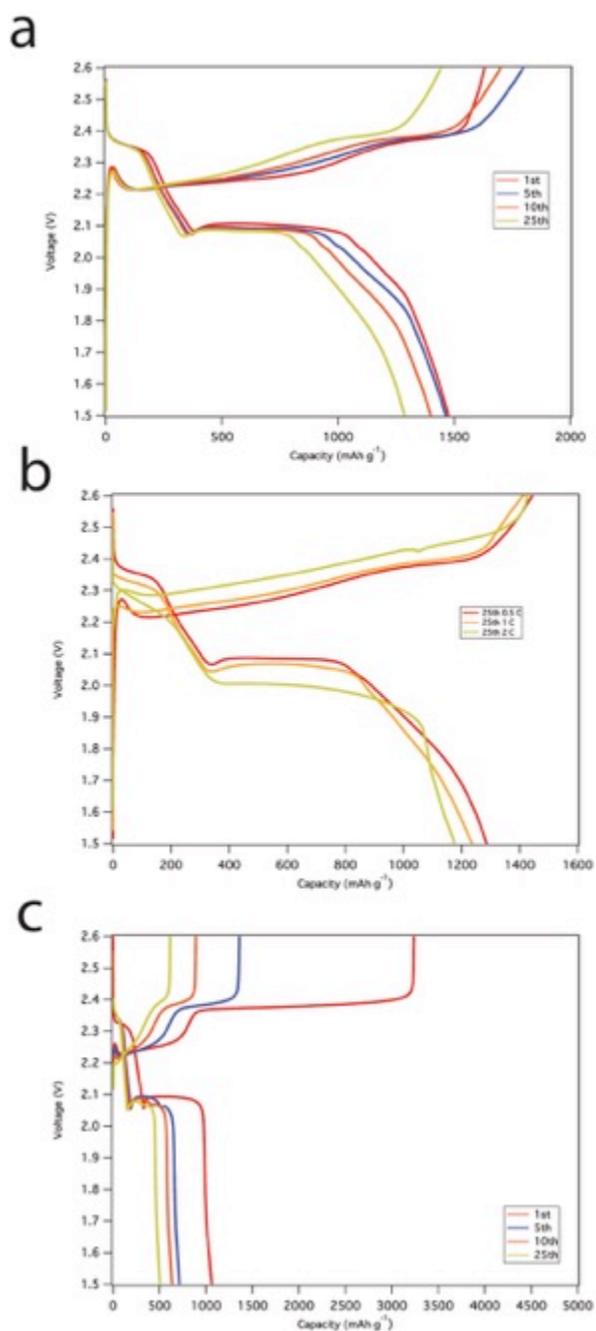

**Supplementary Figure S16: Voltage profiles of Li-S cells with clip coated separator and ISC.** (a) Discharge-charge voltage profiles of the pristine Li anode Li-S cell with the clip coated separator and ISC for various cycles at 0.5 C. (b) Discharge-charge voltage profiles of pristine Li anode Li-S cell with the clip coated separator and ISC at 25$^{th}$ cycle for various C rates. (c) Discharge-charge voltage profiles of pristine Li anode Li-S cell with the pristine separator and ISC at 0.5 C for various cycles.

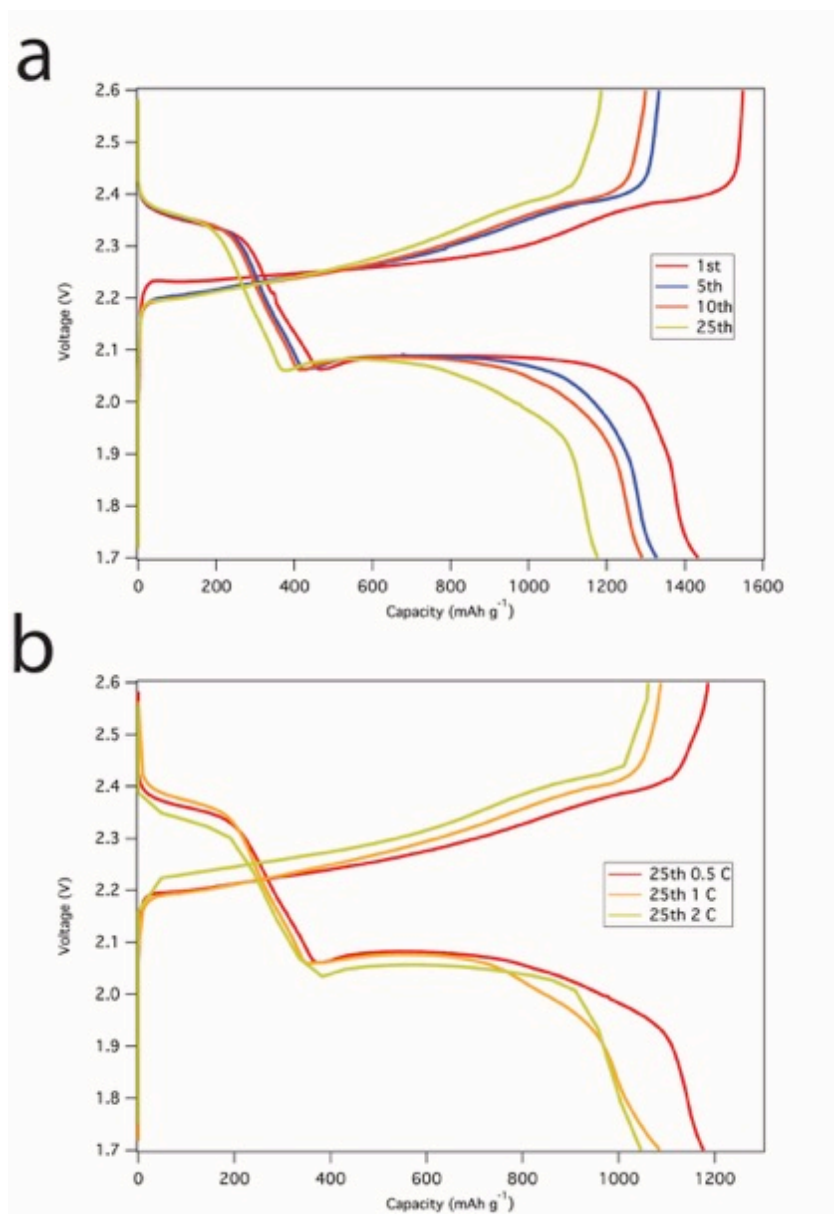

**Supplementary Figure S17: Cycling performance and voltage profiles of Li-S cells with clip coated separator, ISC, and 0.05M LiNO$_3$ in the electrolyte.** (a) Discharge-charge voltage profiles of clip coated separator Li-S cell with 0.05M LiNO$_3$ added in the electrolyte and with ISC for various cycles at 0.5 C. (b) Discharge-charge voltage profiles of the clip coated separator Li-S cell with 0.05M LiNO$_3$ in the electrolyte and with ISC at 25$^{th}$ cycle at various C rates.

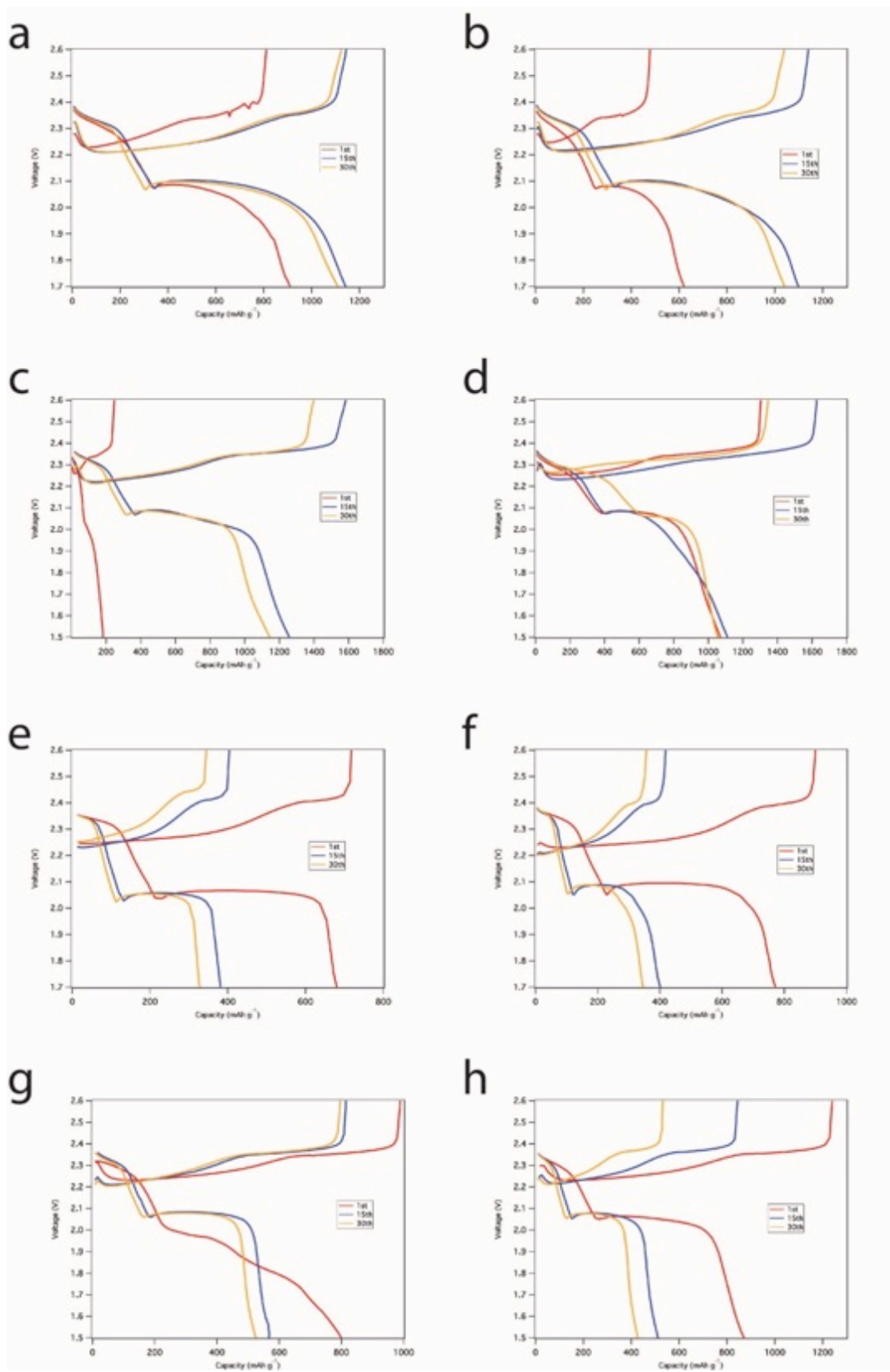

**Supplementary Figure S18: Voltage profiles of Li-S cells with clip coated separator and VISC.** (a) Discharge-charge voltage profiles of clip coated separator Li-S cell with VISC and 0.3M LiNO$_3$ added in the electrolyte for various cycles at 0.5 C. (b) Discharge-charge voltage profiles of clip coated separator Li-S cell with VISC and 0.3M LiNO$_3$ added in the electrolyte for various cycles at 0.2 C. (c) Discharge-charge voltage profiles of clip coated separator Li-S cell with VISC for various cycles at 0.5 C. (d) Discharge-charge voltage profiles of clip coated separator Li-S cell with VISC for various cycles at 0.2 C. (e) Discharge-charge voltage profiles of pristine separator Li-S cell with VISC and 0.3M LiNO$_3$ added in the electrolyte for various cycles at 0.5 C. (f) Discharge-charge voltage profiles of pristine separator Li-S cell with VISC and 0.3M LiNO$_3$ added in the electrolyte for various cycles at 0.2 C. (g) Discharge-charge voltage profiles of pristine separator Li-S cell with VISC for various cycles at 0.5 C. (f) Discharge-charge voltage profiles of pristine separator Li-S cell with VISC for various cycles at 0.2 C.

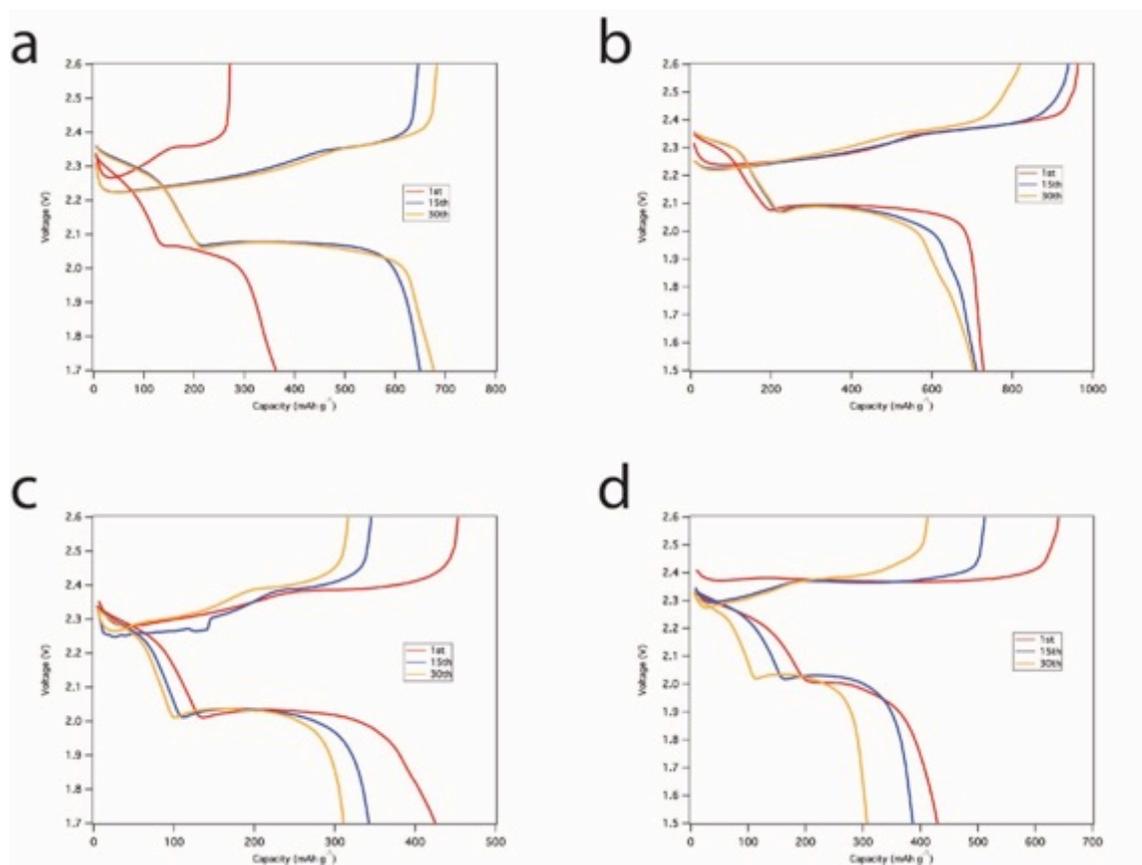

**Supplementary Figure S19: Voltage profiles of Li-S cells with clip coated separator and BMSC.** (a) Discharge-charge voltage profiles of clip coated separator Li-S cell with BMSC and 0.3M LiNO$_3$ added in the electrolyte for various cycles at 0.2 C. (b) Discharge-charge voltage profiles of clip coated separator Li-S cell with BMSC for various cycles at 0.2 C. (c) Discharge-charge voltage profiles of pristine separator Li-S cell with BMSC and 0.3M LiNO$_3$ added in the electrolyte for various cycles at 0.2 C. (d) Discharge-charge voltage profiles of pristine separator Li-S cell with BMSC for various cycles at 0.2 C.

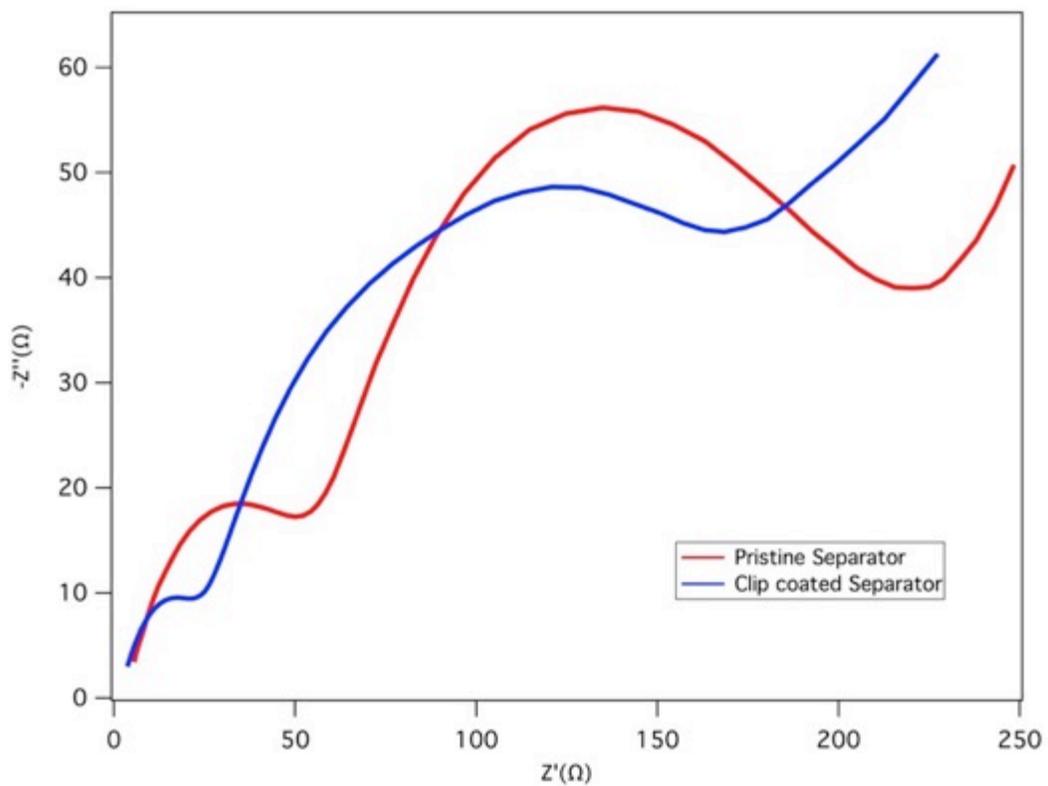

**Supplementary Figure S20:** AC impedance spectroscopy analysis of the clip coated separator and pristine separator Li-S cells with ISC.

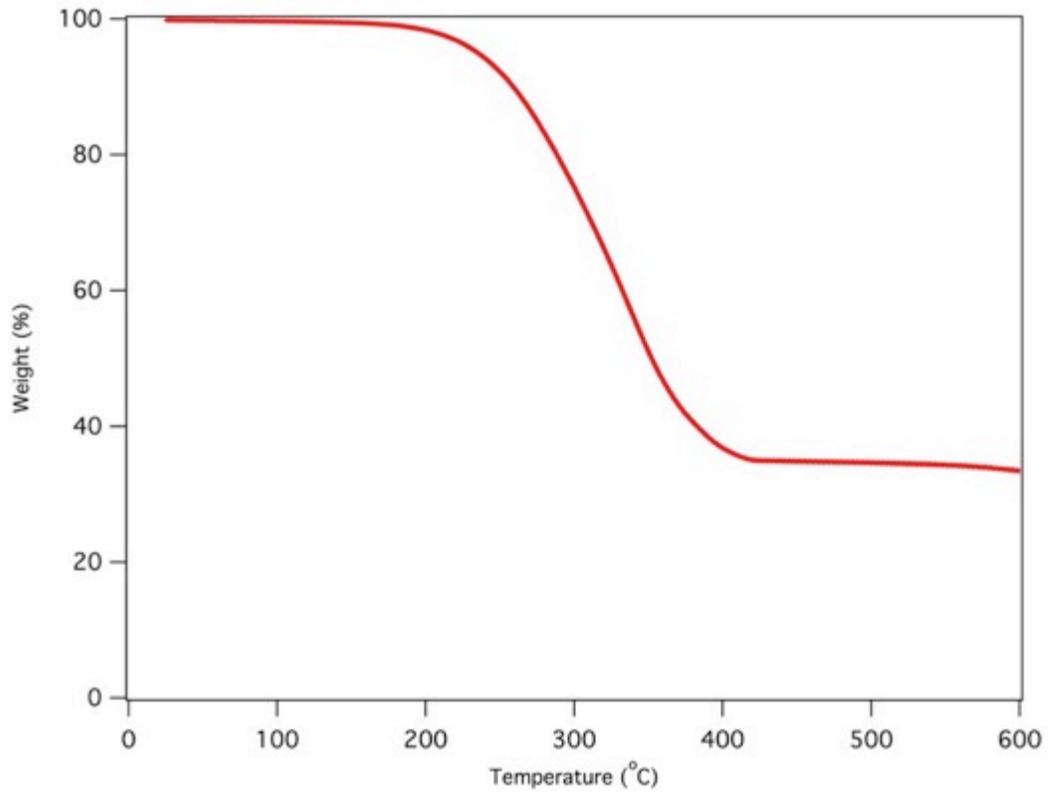

**Supplementary Figure S21: Thermogravimetric analysis of sulfur infused in KB composite in a N₂ gaseous atmosphere with a heating rate of 10 °C min⁻¹, exhibiting 66 wt% sulfur content.**

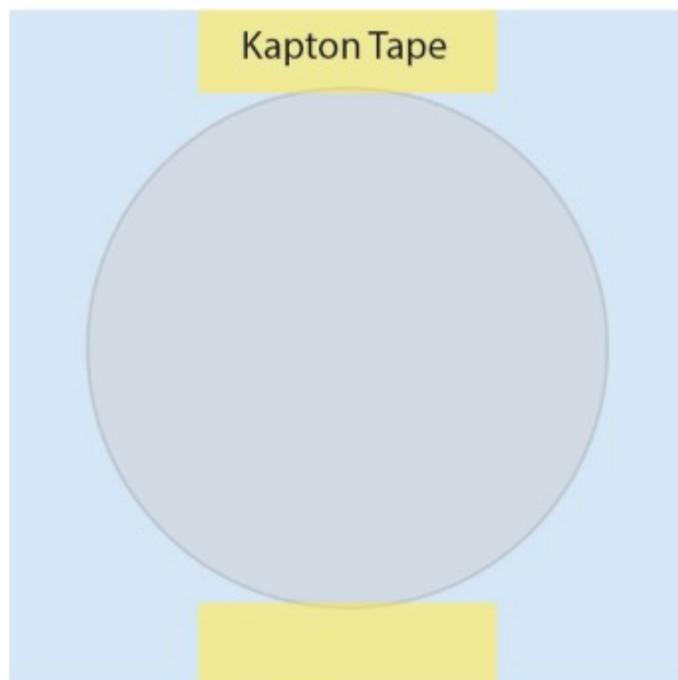 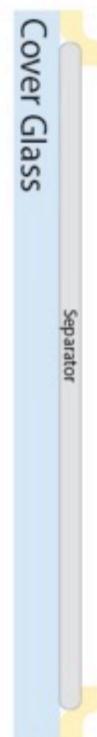

**Supplementary Figure S22: Schematic illustration of as-prepared separator for the LBS and LBSDC coating.**